\pdfoutput=1
\documentclass[times]{speauth}
\usepackage{moreverb}
\usepackage{dcolumn}
\usepackage{listings}
\usepackage{alltt}
\usepackage{setspace}

\usepackage[colorlinks,bookmarksopen,bookmarksnumbered,citecolor=red,urlcolor=red]{hyperref}
\usepackage{xcolor,colortbl,graphicx,amsmath,amsthm,amssymb,subfig,subfloat,verbatim,url,todonotes} 

\newcommand\BibTeX{{\rmfamily B\kern-.05em \textsc{i\kern-.025em b}\kern-.08em
T\kern-.1667em\lower.7ex\hbox{E}\kern-.125emX}}

\newcommand{\addth}{$^{\mathrm{th}}$ }

\begin{document}

\graphicspath{{../epfor-cpp/extras/}{./figures/}}
 
\runningheads{D.~Lemire and L. Boytsov}{Decoding billions of integers per second through vectorization}

\title{Decoding billions of integers per second through vectorization}

\author{D.~Lemire\affil{1}\corrauth,  L. Boytsov\affil{2}}

\address{\affilnum{1}LICEF Research Center, TELUQ, Montreal, QC, Canada\break
\affilnum{2}Carnegie Mellon University, Pittsburgh, Pennsylvania, USA\break
}

\corraddr{LICEF Research Center, TELUQ, Universit\'e du Qu\'ebec, 5800 Saint-Denis,  Montreal (Quebec) H2S 3L5 Canada.}

\cgsn{Natural Sciences and Engineering Research Council of Canada}{261437}
\begin{abstract}
In many important applications---such as search engines and relational database systems---data is stored in the form of arrays of integers.
Encoding and, most importantly, decoding of these arrays
consumes considerable CPU~time.
Therefore,  substantial effort has been made to reduce costs associated
with compression and decompression. 
In particular,  researchers have exploited the superscalar nature of modern processors and SIMD instructions. Nevertheless, we introduce a novel  vectorized scheme called \mbox{SIMD-BP128$^{\star}$} that improves over previously proposed vectorized approaches.
It is nearly twice 
as fast as  the previously fastest schemes on desktop processors (varint-G8IU and PFOR).
At the same time, \mbox{SIMD-BP128$^{\star}$} saves up to 2~bits  per integer.
For even better compression, we propose another new vectorized  
 scheme (SIMD-FastPFOR) that has a compression ratio within 10\% of a state-of-the-art scheme (Simple-8b) 
while being two times faster during decoding.

\end{abstract}


\maketitle

\section{Introduction}

Computer memory is a hierarchy of storage devices 
that range from slow and inexpensive (disk or tape) to fast but expensive (registers or CPU~cache). 
In many situations, application performance is inhibited by access to slower storage devices,
at lower levels of the hierarchy.
Previously, only disks and tapes were considered to be slow devices. 
Consequently, application developers tended to optimize only disk and/or tape I/O. 
Nowadays, CPUs have become so fast that access to main memory is a limiting factor for many workloads~\cite{springerlink:10.1007/3-540-44681-8_63,drepper-memory,Manegold:2009:DAE:1687553.1687618,Zukowskithesis,Harizopoulos:2006:PTR:1182635.1164170}:  data compression can significantly improve query performance by reducing the main-memory bandwidth requirements.

Data compression helps to load and keep more of the data into a faster storage. 
Hence, high speed compression schemes can improve the performances of database systems~\cite{Westmann:2000:IPC:362084.362137,1142548,Buttcher:2007:ICG:1321440.1321546}
and text retrieval engines~\cite{1034897,yan2009inverted,Yandex:2010,Stepanov:2011:SDP:2063576.2063627,DeanOfficialplusslides:2009:CBL:1498759.1498761}.

We focus on compression techniques for 32-bit integer sequences. 
It is best if most of the integers are small, 
because we can save space by representing small integers more compactly, i.e., using short codes.
Assume, for example, that none of the values is larger than 255. 
Then we can encode each integer using one byte,
thus, achieving  a compression ratio of 4: an integer uses 4 bytes in the uncompressed format.

In relational database systems, column values are  transformed into integer values by dictionary coding~\cite{Lemke:2010:SUQ:1881923.1881936,1559877,Poess:2003:DCO:1315451.1315531,click2012,raman2006wring}.
To improve compressibility,
we may map the most frequent values to the smallest integers~\cite{rlewithsorting}.
In text retrieval systems, word occurrences are commonly represented by sorted lists of integer document identifiers, 
also known as posting lists. 
These identifiers are converted to small integer numbers through data differencing.
Other database indexes can also be stored similarly~\cite{1646158}.
\begin{figure}
\centering
\subfloat[encoding]{%
\textrm{array}$\to$ 
\begin{tabular}{|c|}\hline
differential coding\\
 (e.g., $\delta_i=x_i-x_{i-1}$)\\
\hline
\end{tabular}
 $\to $
\begin{tabular}{|c|}\hline
compression\\
 (e.g., SIMD-BP128)\\
\hline
\end{tabular} 
 $\to \mathrm{compressed}$
} 

\subfloat[decoding]{%
$\mathrm{compressed}
\to $
\begin{tabular}{|c|}\hline
decompression \\
 (e.g., SIMD-BP128)\\
\hline
\end{tabular}
$\to$ 
\begin{tabular}{|c|}\hline
differential decoding\\
 (e.g., $x_i=\delta_i+x_{i-1}$)\\
\hline
\end{tabular}
$\to \mathrm{array}$
}
\caption{\label{fig:generalframework}Encoding and decoding of integer arrays using differential coding and an integer compression algorithm}
\end{figure}

A mainstream approach to data differencing is differential coding (see Fig.~\ref{fig:generalframework}). Instead of storing the original array of sorted integers  ($x_1, x_2,  \ldots$ with $x_i\leq x_{i+1}$ for all $i$), 
we keep only the difference between successive elements together with the initial value: ($ x_1, \delta_2 = x_2 - x_1, \delta_3 = x_3 - x_2, \ldots$). 
The differences (or deltas) are non-negative integers that are typically much smaller than the original integers. Therefore,
they can be compressed more efficiently.
 We can then reconstruct the original arrays by computing prefix sums ($x_j = x_1 + \sum_{i = 2}^j \delta_j$).
  Differential coding is also known as delta coding~\cite{raman2006wring,1247525,1453912}, not to be confused with Elias delta coding (\S~\ref{sec:elias}).
 A possible downside of differential coding is that random
 access to an integer located at a given index may require summing up
 several deltas: if needed, we can alleviate this problem by partitioning large arrays into smaller ones.

An engineer might be tempted to compress the result using  generic compression tools such as LZO, Google Snappy, FastLZ, LZ4 or gzip.  Yet this might be ill-advised. Our fastest schemes are an order of magnitude faster than a fast generic library like Snappy, while compressing better (\mbox{see \S~\ref{sec:realistic}}).

Instead, it might be preferable to compress these arrays of integers using specialized schemes based on Single-Instruction, Multiple-Data (SIMD) operations. 
Stepanov et al.~\cite{Stepanov:2011:SDP:2063576.2063627} 
reported that their SIMD-based  varint-G8IU algorithm outperformed 
the classic variable byte coding method (see \S~\ref{sec:vb}) by 300\%.
They also showed that use of SIMD instructions allows one to improve performance of decoding algorithms by more than 50\%.

In Table~\ref{table:overall}, we report the speeds of the fastest decoding algorithms reported in the literature on desktop processors.
These numbers cannot be directly compared since hardware, compilers, benchmarking methodology, and data sets differ. 
However, one can gather that varint-G8IU---which can 
be viewed as an improvement on the Group Varint Encoding~\cite{DeanOfficialplusslides:2009:CBL:1498759.1498761} (varint-GB) 
used by Google---is, probably, the fastest  method (except for our new schemes) in the literature. 
According to our own experimental evaluation (see Tables~\ref{table:synth},~\ref{table:aggrealistics} and Fig.~\ref{fig:unaggregated}),
varint-G8IU is indeed one of the most efficient methods,
but there are previously published schemes 
that offer similar or even slightly better performance 
such as PFOR~\cite{1617427}. 
We, in turn, were able to further surpass the decoding speed of 
varint-G8IU by a factor of two while improving the compression ratio. 

We report our own speed in
a conservative manner: (1)~our timings are based on the wall-clock time and not the commonly used CPU time, (2)~our timings incorporate all of the decoding operations including the computation of the prefix sum whereas this is sometimes omitted by other authors~\cite{Silvestri:2010:VEC:1871437.1871592}, (3)~we report a speed of 2300~million integers per second (mis) achievable for realistic data sets, while higher speed is possible (e.g., we report a speed of 2500\,mis on some realistic data  and 2800\,mis on some synthetic data).

Another observation we can make from Table~\ref{table:overall} is that not all authors have chosen to make explicit use of SIMD instructions. While there are has been several variations on PFOR~\cite{1617427} such as NewPFD and OptPFD~\cite{yan2009inverted}, we introduce for the first time a variation designed to exploit the vectorization instructions available since the introduction of the Pentium~4 and the Streaming SIMD Extensions~2 (henceforth SSE2). Our experimental results indicate that such vectorization is desirable: our SIMD-FastPFOR$^{\star}$ scheme  surpasses the decoding speed of PFOR by at least 30\% while offering a superior compression ratio (10\%).  In some instances, SIMD-FastPFOR$^{\star}$ is twice as fast as the original PFOR\@.

For most schemes,
the prefix sum computation is so fast as to represent 20\% or less of the running time. However, because our novel schemes are much
faster, the prefix sum can account for the majority of the running time.

Hence, we had to experiment with faster alternatives. We find that a
 vectorized prefix sum using SIMD instructions can be twice as fast. Without vectorized differential coding, we were unable to reach a speed of two billion integers per second.

 In a sense, the speed gains we have
achieved are a direct application of advanced hardware
instructions to the problem of integer coding (specifically SSE2 introduced in 2001). Nevertheless, it is
instructive to show how this is done, and to quantify the benefits
that accrue.

\begin{table}
\caption{\label{table:overall}Recent best decoding speeds in millions of 32-bit integers per second (mis) reported by authors for integer compression on realistic data. We indicate whether the authors made explicit use of SIMD instructions.
Results are not directly comparable but they illustrate the evolution of performance.
}\setlength{\tabcolsep}{4pt}
\small\begin{tabular}{lrclll}
 & Speed & Cycles/int & Fastest scheme & Processor & SIMD \\[1.5ex] 
 \textbf{this paper} & \textbf{2300} & \textbf{1.5} & \textbf{SIMD-BP128$^{\star}$} &  Core~i7 (3.4\,GHz) & SSE2 \\
Stepanov et al.\ (2011)~\cite{Stepanov:2011:SDP:2063576.2063627} & 1512  & 2.2 & varint-G8IU &   Xeon (3.3\,GHz) & SSSE3\\
Anh and Moffat (2010)~\cite{Anh:2010:ICU:1712666.1712668} &    1030 &  2.3 & binary packing &  Xeon (2.33\,GHz) & no\\
Silvestri and Venturini (2010)~\cite{Silvestri:2010:VEC:1871437.1871592} & 835 & --- & VSEncoding &  Xeon  & no\\
Yan et al.\ (2009)~\cite{yan2009inverted} & 
1120 & 2.4  & NewPFD &  Core~2 (2.66\,GHz) & no\\
Zhang et al.\ (2008)~\cite{1367550} & 890 & 3.6 & PFOR2008 & Pentium~4  (3.2\,GHz)  & no\\
Zukowski et al. (2006)~\cite[\S~5]{1617427} & 1024 & 2.9 & PFOR & Pentium~4 (3\,GHz)  & no \\
\end{tabular}
\end{table}

\section{Related work}

Some of the earliest integer compression techniques are Golomb coding~\cite{323905}, Rice coding~\cite{1090789}, as well as Elias gamma and delta coding~\cite{elias1975universal}. In recent years, several faster techniques have been added such as the Simple family, binary packing, and patched coding.
We briefly review them.

Because we work with unsigned integers, we make use of two representations: binary and unary. 
In both systems numbers are represented using only two digits: 0 and 1.
The binary notation is a standard positional base-2 system 
(e.g.,  $1\to 1$, $2\to  10$, $3\to 11$). 
Given a positive integer $x$, the binary notation requires  $\lceil \log_2 (x + 1)\rceil$~bits. 
Computers commonly store
unsigned integers in the binary notation using a fixed number of bits by adding leading zeros: e.g., $3$ is written as  $00000011$ using 8~bits.
In unary notation, we represent a number $x$ as a sequence of $x-1$~digits~0 followed by the digit 1 (e.g.,  $1\to 1$, $2\to  01$, $3\to 001$)~\cite{buttcher2010information}. 
If the number $x$ can be zero, we can store $x+1$ instead.

\subsection{Golomb and Rice coding}

In Golomb coding~\cite{323905}, given a fixed parameter $b$ and a positive integer $v$ to be compressed, the quotient  $\lfloor v /b \rfloor$ is coded  in unary. The remainder $r = v \bmod b$ is stored using the usual binary notation with no more than $\lceil \log_2 b \rceil$~bits.
 If $v$ can be zero, we can code $v+1$ instead.
  When $b$ is chosen to be a power of two, the resulting algorithm is called Rice coding~\cite{1090789}.
The parameter $b$ can be chosen optimally by assuming some that the integers follow a known distribution~\cite{323905}.

Unfortunately,
Golomb and Rice coding are much slower than a simple scheme such as Variable Byte~\cite{1034897,yan2009inverted,Transier:2010:EBA:1877766.1877768} (see \S~\ref{sec:vb}) which, itself, falls short of our goal of  decoding billions of integers per second  (\mbox{see \S~\ref{sec:synth}--\ref{sec:realistic}}).

\subsection{Interpolative coding}

If speed is not an issue but high compression over sorted arrays is desired,  
interpolative coding~\cite{moffat2000binary} might be appealing. 
In this scheme, we first store the lowest and the highest value, $x_1$ and $x_n$, e.g., in a uncompressed form.
Then a  value in-between is stored in a binary form,
using the fact this value must be in the range $(x_1, x_n)$. 
For example, if $x_1=16$ and $x_n=31$, 
we know that for any value $x$ in between, the difference $x - x_1$ is from 0 to 15.
Hence, we can encode this difference using only 4 bits.
The technique is then repeated recursively.
Unfortunately, it is slower than Golomb coding~\cite{1034897,yan2009inverted}.

\subsection{Elias gamma and delta coding}

\label{sec:elias}

An Elias gamma code~\cite{elias1975universal,buttcher2010information,Walder:2012:FDA:2051366.2051431} consists of two parts. The first part encodes in unary notation  the minimum number of bits necessary to store the positive integer in the binary notation ($\lceil \log_2 (x+1)\rceil$).
The second part represents the integer in binary notation less the most significant digit. If the integer is equal to  one, the second part is empty (e.g., $1\to 1$, $2\to 01\,0$, $3\to 01\,1$, $4\to 001\,00$, $5\to 001\,01$). If integers can be zero, we can  code their values incremented by one.

As numbers become large,  gamma codes become inefficient.
For better compression, Elias delta codes encode the first part (the number $\lceil \log_2 (x+1)\rceil$) using  the Elias gamma code, while the second part is coded in the binary notation. For example, to code the number 8 using the Elias delta code, we must first store $4= \lceil \log_2 (8+1)\rceil$ as a gamma code ({\color{blue}$ 001\,00$}) and then we can store all but the most significant bit of the number 8 in the binary notation ({\color{red}$000$}). The net result is {\color{blue}$  001\,00$}~{\color{red}$000$}.

However, Variable~Byte is twice as fast as Elias gamma and delta coding~\cite{Silvestri:2010:VEC:1871437.1871592}. Hence, like Golomb coding, gamma coding falls short of our objective of compressing billions of integers per second.

\subsubsection{$k$-gamma}

Schlegel et al.~\cite{Schlegel:2010:FIC:1869389.1869394} proposed a version
of Elias gamma coding better suited to current processors.
To ease vectorization, the data is stored in blocks of $k$~integers using the same number of bits where $k\in\{2,4\}$. (This approach is similar to binary packing described in \S~\ref{sec:binpacking}.) As with regular gamma coding, we use unary codes to store this number of bits though we only have one such number for $k$~integers. 

The binary part of the gamma codes are  stored using the same vectorized layout as in \S~\ref{sec:fast-bit-unpacking} (known as vertical or interleaved). During decompression, we decode integer in groups of $k$~integers.
For each group we first retrieve the binary length from a gamma code. 
Then, we decode group elements using a sequence of mask-and-shift operations similar to the fast bit unpacking technique described in \S~\ref{sec:fast-bit-unpacking}.
This step does not require branching.

Schlegel et al.\ report best decoding speeds of $\approx $550\,mis ($\approx $2100\,MB/s) on synthetic data using an Intel Core i7-920 processor (2.67\,GHz).
These results fall short of our objective to compress billions of integers per second.

\subsection{Variable Byte and byte-oriented encodings}
\label{sec:vb}

Variable Byte is a popular technique~\cite{Cutting:1989:ODI:96749.98245} that is known under many names (v-byte, variable-byte~\cite{williams1999compressing}, var-byte, vbyte~\cite{buttcher2010information}, varint, VInt, VB~\cite{Stepanov:2011:SDP:2063576.2063627}
 or Escaping~\cite{Transier:2010:EBA:1877766.1877768}).
 To our knowledge, it was first described by Thiel and Heaps in 1972~\cite{thiel1972program}.
 Variable Byte codes  the data in units of bytes:
it uses the lower-order seven bits to store the data, 
while the eighth bit is used as an implicit indicator of a code length. 
Namely, the eighth bit is equal to 1 only for the last byte of a sequence that encodes an integer. 
For example:\begin{itemize}
\item  Integers in $[0,2^7)$ are written using one byte: The first 7~bits are used to store the binary representation of the integer and the eighth bit is set to 1.
\item Integers in $[2^7,2^{14})$ are written using two bytes, the eighth bit of the first byte is set to 0 whereas the eighth bit of the second byte is set to 1. The remaining 14~bits are used to store the binary representation of the integer.   
\end{itemize}
For a concrete example, consider the number 200. It is written as 11001000 in the binary notation. Variable~Byte would code it using 16~bits  as  {\color{red}1}000000{\color{blue}1}~{\color{red}0}{\color{blue}1001000}. 

When decoding, bytes are read one after the other: we  discard the eighth bit if it is zero, 
and we output a new integer whenever the eighth bit is one. 

Though Variable Byte rarely compresses data optimally, it is reasonably efficient.  
In our tests, Variable Byte encodes data  three~times faster than most alternatives. Moreover, when the data is not highly compressible, it can match the compression ratios of more parsimonious schemes.

Stepanov et al.~\cite{Stepanov:2011:SDP:2063576.2063627} generalize Variable~Byte into a family of byte-oriented encodings. Their main characteristic is that each encoded byte contains bits from only one integer. However, whereas Variable~Byte uses one bit per byte as descriptor, alternative schemes can use other arrangements. For example, varint-G8IU~\cite{Stepanov:2011:SDP:2063576.2063627} and Group Varint~\cite{DeanOfficialplusslides:2009:CBL:1498759.1498761} (henceforth varint-GB) regroup all descriptors in a single byte. Such alternative layouts make easier the simultaneous decoding of several integers. 
A similar approach to placing descriptors in a single control word was used
to accelerate a variant of the Lempel-Ziv algorithm~\cite{Williams:1991}.
 
For example, varint-GB uses a single byte to describe 4~integers, dedicating 2~bits per integer. The scheme is better explained by an example. Suppose that we want to store the integers $2^{15}$, $2^{23}$, $2^{7}$, and $2^{31}$. In the usual binary notation, we would use 2, 3, 1 and 4~bytes, respectively. We can store the sequence as 2, 3, 1, 4 as 1, 2, 0,
3 
if we assume that each number is encoded using a non-zero number of bytes.
Each one of these 4~integers can be written using 2~bits (as they are in \{0,1,2,3\}). We can pack them into a single byte containing the bits 01,10,00, and 11. Following this byte,  we write the integer values using $2+3+1+4=10$~bytes.

Whereas varint-GB codes a fixed number of integers (4) using a single descriptor, 
varint-G8IU uses a single descriptor for a group of 8~bytes, which represent compressed integers. Each 8-byte group may store from 2 to 8~integers.
A single-byte descriptor is placed immediately before this 8-byte group.
Each bit in the descriptor represents a single data byte.
Whenever a descriptor bit is set to~0, then the corresponding byte is the end of an integer. 
This is symmetrical to the Variable Byte scheme described above,
where the descriptor bit value~1 denotes the last byte of an integer code.

In the example we used for varint-GB, we could only store the first 3~integers ($2^{15}$, $2^{23}$, $2^{7}$) into a single 8-byte group, 
because storing all 4~integers would require 10 bytes.
These integers use 2, 3, and 1 bytes, respectively, whereas the descriptor byte is equal to 11{\color{blue}0}{\color{red}011}{\color{blue}01}  (in the binary notation). 
The first two bits (\textcolor{blue}{01}) of the descriptor tell us that the first integer uses 2~bytes.
The next three bits (\textcolor{red}{011}) indicate that the second integer requires 3~bytes.
Because the third integer uses a single byte, the next (sixth) bit of the descriptor would be {\color{blue}0}. 
In this model, the last two bytes cannot be used and, thus, we would set the last two bits to 1. 

\begin{figure}\centering
\includegraphics[width=0.8\textwidth]{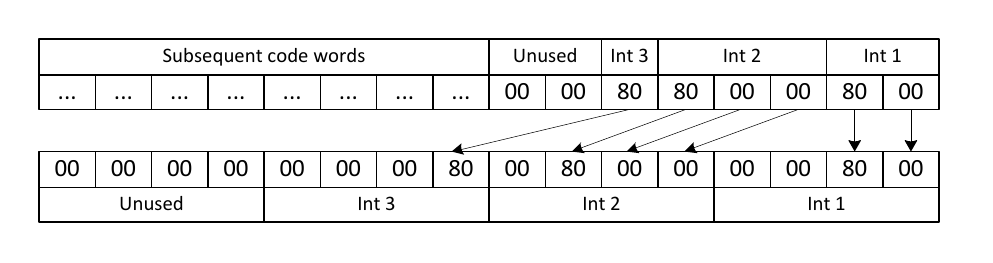}
\caption{\label{fig:varint-g8iu} Example of simultaneous decoding of 3~integers 
in the scheme varint-G8IU using the shuffle instruction.
The integers $2^{15}$, $2^{23}$, $2^{7}$ are packed into the 8-byte block with 2~bytes being unused.
Byte values are given by \emph{hexadecimal} numbers.
The target 16-byte buffer bytes are either copied from the source 16-byte buffer or are filled with zeros.
Arrows indicate which bytes of the source buffer are copied to the target buffer as well
 as their location in the source and target buffers. 
}
\end{figure}

On most recent x86 processors,
integers packed  with \mbox{varint-G8IU} can be efficiently decoded using the SSSE3 (Supplemental Streaming SIMD Extensions~3) \emph{shuffle} instruction:
\texttt{pshufb}.
This assembly operation selectively copies byte elements of a 16-element vector to specified locations 
of the target 16-element buffer
and replaces selected elements with zeros.

The name ``shuffle'' is a misnomer, because certain source bytes can be omitted,
while others may be copied multiple times to a number of different locations.
The operation takes two 16~element vectors (of $16\times 8 = 128$~bits each): the first vector contains the bytes to be shuffled into an output vector whereas the second vector serves as a \emph{shuffle mask}. 
Each byte in the shuffle mask determines which value will go in the corresponding location in the output vector. If the last bit is set (that is, if the value of the byte is larger than 127), the  target byte is zeroed.
For example, if the shuffle mask contains the byte values $127, 127, \ldots, 127$, 
then the output vector will contain only zeros.
Otherwise, the first 4~bits of the $i$\addth mask element determine the index of the byte that should
be copied to the target byte $i$. For example, if the shuffle mask contains the byte values $0, 1, 2, \ldots, 15$, then the bytes are simply copied in their original locations.

In Fig.~\ref{fig:varint-g8iu},
we illustrate one step of the decoding algorithm for varint-G8IU\@.
We assume that the descriptor byte, which encodes the 3~numbers of bytes (2, 3, 1) required to store the   
3~integers ($2^{15}$, $2^{23}$, $2^{7}$), is already retrieved.
The value of the descriptor byte was used to obtain a proper shuffle mask for \texttt{pshufb}.
This mask 
 defines a 
 hardcoded
sequence of operations 
that copy bytes from the source to the target buffer or fill selected bytes of the target 
buffer with zeros.
All these byte operations are carried out in parallel in the following manner
(byte numeration starts from \emph{zero}):
\begin{itemize}
\item The first integer uses  2~bytes, which are both copied to bytes 0--1 of the target buffer
Bytes 2--3 of the target buffer are zeroed.
\item Likewise, we copy bytes 2--4 of the source buffer to bytes 4--6 of the target buffer.
Byte 7 of the target buffer is zeroed.
\item The last integer uses only one byte 5: we copy the value of this byte to byte 8 and 
zero bytes 9--11.
\item The bytes 12--15 of the target buffer are currently unused and will be filled out by subsequent
decoding steps. In the current step, we may fill them with arbitrary values, e.g., zeros.
\end{itemize}

We do not know whether Google implemented varint-GB  using SIMD instructions~\cite{DeanOfficialplusslides:2009:CBL:1498759.1498761}. However,
Schlegel et al.~\cite{Schlegel:2010:FIC:1869389.1869394} and Popov~\cite{Yandex:2010}
described the application of the \texttt{pshufb} instruction to accelerate decoding of a varint-GB scheme (which Schlegel et al. called \emph{4-wise null suppression}).

Stepanov et al.~\cite{Stepanov:2011:SDP:2063576.2063627}  found varint-G8IU to compress slightly better than a SIMD-based varint-GB while being up to 20\% faster. Compared to the common Variable~Byte, varint-G8IU had a slightly worse compression ratio (up to 10\%), but it is  2--3~times faster.

\subsection{The Simple family}

Whereas Variable Byte takes a fixed input length (a single integer) and produces a variable-length output (1, 2, 3 or more bytes), at each step 
the Simple family outputs a fixed number of bits, but processes a variable number of integers, similar to varint-G8IU\@. 
However, unlike varint-G8IU, schemes from the Simple family are not byte-oriented. Therefore, they may fare better on highly compressible arrays (e.g., they could compress a sequence of numbers in $\{0,1\}$ to $\approx$1\,bit/int).

The most competitive Simple scheme on 64-bit processors is Simple-8b~\cite{Anh:2010:ICU:1712666.1712668}. It outputs 64-bit words. The first 4~bits of every 64-bit word is a selector that indicates an encoding mode. 
 The remaining 60~bits are employed to keep data. Each integer is stored using the same number of bits $b$.
Simple-8b has 2 schemes to encode long sequences of zeros and 14 schemes to encode positive integers.
For example:
\begin{itemize}
\item Selector values 0 or 1 represent sequences containing 240 and 120~zeros,
respectively. In this instance the 60~data bits are ignored. 
\item The selector value 2 corresponds to $b=1$. This allows us to store 60~integers having values in \{0,1\}, which are packed in the data bits.
\item The selector value 3 corresponds to $b=2$ and allows one to pack 30~integers having values in $[0,3]$ in the data bits. 
\end{itemize}
And so on (see Table~\ref{table:simple8b}):  the larger is the value of the selector, the larger is $b$, and the fewer
integers one can fit in 60 data bits. 
During coding, we try successively the selectors starting with value 0. That is, we greedily try to fit as many integers as possible in the next 64-bit word.

\begin{table}
\caption{Encoding mode for Simple-8b scheme. Between 1 and 240~integers are coded with one 64-bit word.\label{table:simple8b}
}
\centering
\begin{tabular}{lrrrrrrrrrrrrrrrr}
\small selector value 	 &\small	 0 	 &\small	 1 	 &\small	  2 	 &\small	 3 	 &\small	 4 	 &\small	 5 	 &\small	 6 	 &\small	 7 	 &\small	 8 	 &\small	 9 	 &\small	 10 	 &\small	 11 	 &\small	 12 	 &\small	 13 	 &\small	 14 	 &\small	 15  \\
\hline
\small   integers coded 	 &\small	  240 	 &\small	  120	 &\small	  60 	 &\small	   30 	 &\small	   20 	 &\small	   15 	 &\small	   12 	 &\small	   10 	 &\small	   8 	 &\small	   7 	 &\small	   6 	 &\small	   5 	 &\small	   4 	 &\small	   3 	 &\small	   2 	 &\small	   1  \\
\hline
\small  bits per integer 	 &\small	  0 	 &\small	  0 	 &\small	  1 	 &\small	  2 	 &\small	  3 	 &\small	  4 	 &\small	  5 	 &\small	  6 	 &\small	  7 	 &\small	  8 	 &\small	  10 	 &\small	  12 	 &\small	  15 	 &\small	  20 	 &\small	  30 	 &\small	  60  \\

\end{tabular}
\end{table}

Other schemes such as Simple-9~\cite{1034897} and Simple-16~\cite{yan2009inverted} use words of 32~bits. (Simple-9 and Simple-16 can also be written as S9 and S16~\cite{yan2009inverted}.) While these schemes may sometimes compress slightly better, they are generally slower. Hence, we omitted them in our experiments. 
Unlike Simple-8b that can encode integers in $[0,2^{60})$, Simple-9 and Simple-16 are restricted to integers in $[0,2^{28})$.

While Simple-8b is not as fast as Variable Byte during encoding, it is still faster than many alternatives. Because the decoding step can be implemented efficiently (with little branching), we also get a good decoding speed while achieving a better compression ratio than Variable Byte.

\subsection{Binary Packing}
\label{sec:binpacking}
Binary packing is closely related to Frame-Of-Reference (FOR) from Goldstein et al.~\cite{655800} and tuple differential coding from Ng and Ravishankar~\cite{ng1997block}. In such techniques, arrays of values are partitioned into blocks (e.g., of 128~integers). 
The range
of values in the blocks is first coded and then all values in the block are written in reference to the range of values: 
for example, if the values in a block are integers in the range $[1000,1127]$, then they can be stored using 7~bits per integer ($\lceil \log_2 (1127+1-1000) \rceil = 7$) as offsets from the number 1000 stored in the binary notation. 
In our approach to binary packing, we assume that integers are small, so we only need to code a bit width $b$ per block (to
represent the range). Then, successive values are stored using $b$~bits per integer using fast bit packing functions.
Anh and Moffat called binary packing  \emph{PackedBinary}~\cite{Anh:2010:ICU:1712666.1712668} whereas Delbru et al.~\cite{Delbru2011} called their 128-integer binary packing  FOR and their 32-integer binary packing AFOR-1.

Binary packing can have a competitive compression ratio. 
In Appendix~\ref{sec:itbinpack}, we derive a general information-theoretic lower bound on the compression ratio of binary packing.

\subsection{Binary Packing with variable-length blocks}

Three factors determine to the storage cost of a given block
in binary packing:
\begin{itemize}
\item the number of bits ($b$) required to store the largest integer value
in the binary notation,
\item the block length ($B$),
\item and a fixed per-block overhead ($\kappa$). 
\end{itemize}
The total storage cost for one block is $b B + \kappa$.
Binary packing uses fixed-length blocks (e.g., $B=32$ or $B=128$).

We can  vary dynamically the length of the blocks
to improve the compression ratio. 
This adds a small overhead to each block because we need to store
not only the corresponding bit width ($b$) but also the block length ($B$).
We then have a conventional optimization problem: we must partition
the array into blocks so that the total storage cost is minimized.
The cost of each block is still given by  $b B + \kappa$, but the block length $B$ may vary from one block to another.

The dynamic selection of block length was first proposed by  Deveaux et al.~\cite{deveaux2007adaptive} who reported compression gains (15--30\%).
They used both a top-down and a bottom-up heuristic.

 Delbru et al.~\cite{Delbru2011} also implemented  two such adaptive solutions, AFOR-2 and AFOR-3. AFOR-2 picks blocks of length 8, 16, 32 whereas AFOR-3 adds a special processing for the case where we have successive integers. To determine the best configuration of blocks, they pick 32 integers and try various configurations (1~block of 32~integers, 2~blocks of 16~integers and so on). 
They keep the configuration minimizing the storage cost.
In effect, they apply  a greedy approach to the storage minimization problem.

Silvestri and Venturini~\cite{Silvestri:2010:VEC:1871437.1871592} proposed two variable-length schemes, and we selected their fastest version (henceforth VSEncoding). VSEncoding optimizes the block length using dynamic programming over blocks  of lengths 1--14, 16, 32. 
That is, given the integer logarithm of every integer in the array,
VSEncoding finds a partition truly minimizing the total storage 
cost.  
We expect VSEncoding to provide a superior compression ratio compared to AFOR-2 and AFOR-3. 

\subsection{Patched coding}
\label{sec:patchedcoding}

Binary packing might sometimes compress poorly. 
 For example, the integers $1,4,255, 4, 3, 12, 101$ can be stored using slightly more than 8~bits per integer with binary packing. However, the same sequence with one large value, e.g., $1,4,255, 4, 3, 12, 4294967295$, is no longer so compressible: at least $32$~bits per integer are required.
 Indeed, 32~bits are required for storing $4294967295$ in the binary notation and all integers use the same bit width under binary packing.

To alleviate this problem, Zukowski et al.~\cite{1617427} proposed \emph{patching}: we use a small bit width $b$, but store exceptions (values greater than or equal to $2^b$) in a separate location. 
They called this approach PFOR\@. (It is sometimes written 
PFD~\cite{Ao:2011:EPL:2002974.2002975}, PFor or PForDelta when used in conjunction with differential coding.) 
We begin with a partition of the input array into 
subarrays that have a fixed maximal size
(e.g., 32\,MB). We call each 
such subarray a \emph{page}.

A single bit width is used for an entire page in PFOR\@. 
To determine the best bit width $b$ during encoding, 
a sample of at most $2^{16}$~integers is created
out of the page.
Then, various bit widths
are tested until  the best compression ratio is achieved.
In practice, to accelerate the computation,  we can construct a histogram, 
recording how many integers have a given integer logarithm ($\lceil \log_2 (x + 1)\rceil$).
 
A page is coded in blocks of 128~integers, with a \emph{separate}  storage array for the exceptions. The blocks are coded  using bit packing. We either pack the integer value itself when the value is regular ($<2^b$), or an integer offset pointing to the next exception in the block of 128~integers when there is one. 
The offset is the difference between the index of the next exception and 
the index of the current exception, minus one. 
For the purpose of bit packing, we store integer values
and offsets in the same array without differentiating them.
For example, consider the following array of integers in the binary notation:
\begin{eqnarray*}
10, 10, 1, 10, 100110, 10, 1, 11, 10, 100000, 10, 110100, \ldots
\end{eqnarray*}
Assume that the bit width  is set to three ($b=3$), then
we have exceptions at positions $4, 9, 11, \ldots$, the offsets are
$9-4 - 1=4, 11-9 - 1=1, \ldots$ In the binary notation we have $4\to 100$ and $1\to 1$, so we would store 
\begin{eqnarray*}
10, 10, 1, 10, {\color{blue}100}, 10, 1, 11, 10, {\color{blue} 1}, 10, \ldots
\end{eqnarray*}
The bit-packed blocks are preceded
by a 32-bit word containing two markers.
The first marker indicates the location of the first exception in the block of 128~integers (4 in our example), 
and the second marker indicates the location of this first exception value in the array of exceptions (exception table). 

Effectively, exception locations are stored using a linked list:  we first read the location of the first exception, then going to this location we find an offset from which we retrieve the location of the next exception, and so on. If the bit width $b$ is too small to store an offset value, that is, if the offset is greater or equal than $2^b$, we have to create a \emph{compulsory} exception in-between. The location of the exception values themselves  are found by incrementing the location of the first exception value in the exception table.

 When there are too many exceptions, these exception tables may overflow and it is necessary to start a new page: Zukowski et al.~\cite{1617427} used pages of 32\,MB.\@ In our own experiments, we partition large arrays into arrays of at most $2^{16}$~integers (see \S~\ref{sec:soft}) so a single page is used in practice.

PFOR~\cite{1617427} does not compress the exception values. In an attempt to improve the compression,  Zhang et al.~\cite{1367550} proposed to store the exception values  using either 8, 16, or 32~bits. We implemented this  approach (henceforth PFOR2008). (See Table~\ref{table:overviewpatched}.)

\subsubsection{NewPFD and OptPFD}

The compression ratios of PFOR and PFOR2008 are relatively modest (see \S~\ref{sec:expe}). For example, we found that they fare  worse than binary packing over blocks of 32~integers (BP32). To get better compression, Yang et al.~\cite{yan2009inverted} proposed two  new schemes called NewPFD and OptPFD.
(NewPFD is sometimes called NewPFOR~\cite{springerlink:10.1007/978-3-642-28997-2_37,springerlink:10.1007/978-3-642-28997-2_35}
whereas OptPFD is also known as OPT-P4D~\cite{Silvestri:2010:VEC:1871437.1871592}.)
Instead of using a single bit width $b$ per page, they use a  bit width per  block of 128~integers. They avoid wasteful  compulsory exceptions: instead of storing exception offsets in the bit packed blocks, they store the first $b$~bits of the exceptional integer value. For example, given the following array
\begin{eqnarray*}
10, 10, 1, 10, 100110, 10, 1, 11, 10, 100000, 10, 110100, \ldots
\end{eqnarray*}
and a bit width of 3 ($b=3$), we would pack 
\begin{eqnarray*}
10, 10, 1, 10, {\color{blue}110}, 10, 1, 11, 10, {\color{blue}0}, 10, {\color{blue}100}, \ldots
\end{eqnarray*}
For each block of 128~integers,  the $32-b$ higher bits of the exception  values ($100, 100, 110, \ldots$  in our example) as well as their locations (e.g., $4, 9, 11, \ldots$ )  are compressed using Simple-16. (We tried replacing Simple-16 with Simple-8b but we found no benefit.)

Each block of 128~coded integers is preceded by a 32-bit word used to store the bit width, the number of exceptions, and the storage requirement of the compressed exception values in 32-bit words. 
NewPFD  determines the bit width $b$ by picking the smallest value of $b$ such that not more than 10\%
of the integers are exceptions. OptPFD picks the value of $b$ maximizing the compression. To accelerate the processing, the bit width is chosen among the integer values  0--16, 20 and 32.

\begin{table}
\caption{\label{table:overviewpatched}Overview of the patched coding schemes: Only PFOR and PFOR2008 generate compulsory exceptions and use a single bit width $b$ per page. Only NewPFD and OptPFD store exceptions on a per block basis. We implemented all schemes with 128~integers per block and a page size of at least $2^{16}$~integers.}\centering
\begin{tabular}{lcccc}
 & compulsory & bit width & exceptions    & compressed exceptions \\
 PFOR~\cite{1617427} & yes & per page & per page & no\\
  PFOR2008~\cite{1367550} & yes & per page & per page & 8, 16, 32 bits \\
   NewPFD/OptPFD~\cite{yan2009inverted} & no & per block  & per block & Simple-16\\
   {FastPFOR} (\S~\ref{sec:fastpfor}) & no & per block & per page & binary packing\\
   {SIMD-FastPFOR} (\S~\ref{sec:fastpfor}) & no & per block & per page & vectorized bin.\ pack.\\
   {SimplePFOR} (\S~\ref{sec:fastpfor}) & no & per block & per page & Simple-8b \\
\end{tabular}
\end{table}

Ao et al.~\cite{Ao:2011:EPL:2002974.2002975} also proposed a version
of PFOR called ParaPFD\@. 
Though it has a worse compression efficiency than NewPFD or PFOR, it is
designed for fast execution on graphical processing units (GPUs).

\section{Fast differential coding and decoding}
\label{sec:fast-delta}

All of the schemes we consider experimentally rely on differential coding over 32-bit unsigned integers.
The computation of the differences (or deltas) is
typically considered a trivial operation which accounts for only a
negligible fraction of the total decoding time.
Consequently, authors do not discuss it. But in our experience,
a straightforward implementation of differential decoding can be four times slower than  the decompression of small integers. 

We have implemented and evaluated two approaches to data differencing:
\begin{enumerate}
\item The standard form of differential coding is simple and requires merely one subtraction per value during encoding ($\delta_i = x_i - x_{i-1} $) and one addition per value during decoding to effectively compute the prefix sum ($x_i = \delta_i + x_{i-1}$). 

\item A vectorized differential coding leaves the first four elements unmodified. 
From each of the remaining elements with index $i$,
we subtract the element with the index $i-4$:   $\delta_i = x_i - x_{i-4}$.
In other words, the original array ($x_1, x_2, \ldots$) 
is converted into ($ x_1, x_2, x_3, x_4, \delta_5 = x_5 - x_1, \delta_6 = x_6 - x_2, \delta_7 = x_7 - x_3, \delta_8 = x_8 - x_4,\ldots$).
An advantage of this approach is that we can compute four differences using a single SIMD operation.
This operation carries out an element-wise subtraction for two four-element \emph{vectors}.
The decoding part is symmetric and involves the addition of the element $x_{i-4}$:  $x_i = \delta_i + x_{i-4}$.
Again, we can use a single SIMD instruction to carry out four additions simultaneously. 
\end{enumerate}

We can get a speed of $\approx$2000\,mis or 1.7~cycles/int with the standard differential decoding (the first approach) by manually unrolling  the loops.
Clearly, it is impossible to decode compressed integers at more than 2~billion integers per second 
if the computation of the prefix sum itself runs at 2~billion integers per second. 
Hence, we implemented a vectorized version of differential coding.
Vectorized differential decoding is much faster ($\approx$5000\,mis vs. $\approx$2000\,mis). However, it comes at a price: 
vectorized deltas are, on average, four times larger which increases the storage cost by up to 2~bits (e.g., see Table~\ref{table:aggrealistics}). 

To prevent memory bandwidth from becoming a bottleneck~\cite{springerlink:10.1007/3-540-44681-8_63,drepper-memory,Manegold:2009:DAE:1687553.1687618,Zukowskithesis,Harizopoulos:2006:PTR:1182635.1164170}, we prefer 
to compute differential coding and decoding  in place.
To this end, we compute deltas in decreasing index order, starting from the largest index. For example, given the integers 1, 4, 13, we first compute
the difference between 13 and 4 which we store in last position (1, 4, {\color{blue}9}), then we compute the difference between 4 and 1 which we store in second position (1, {\color{blue}3}, 9).
In contrast, the differential decoding proceeds in the increasing index order, 
starting from the beginning of the array. 
Starting from 1, 3, 9, we first add 3 and 4 which we store in the second position (1, {\color{blue} 4}, 9), then we add 4 and 9 which we store in the last position (1,  4, {\color{blue} 9}).
Further, our  implementation requires 
two passes: one pass to reconstruct the deltas from their compressed format and another pass to compute the prefix sum  (\S~\ref{sec:soft}). 
To improve data locality and reduce cache misses, 
arrays containing more than $2^{16}$~integers ($2^{16} \times 4\mathrm{\,B}=256\mathrm{\,KB}$) are broken down into smaller arrays
and each array is decompressed independently. 
Experiments with synthetic data have shown that reducing cache misses by breaking down arrays can lead to 
nearly
a significant improvement
 in decoding speed for some schemes without degrading the compression efficiency.

\section{Fast bit unpacking}
\label{sec:fast-bit-unpacking}

Bit packing is a process of encoding small integers in $[0,2^b)$ using $b$ bits each: 
$b$ can be arbitrary and not just 8, 16, 32 or 64. Each number is written using a string of exactly $b$~bits.
Bit strings of fixed size $b$ are concatenated together into a single bit string,
which can span several 32-bit words.
If some integer is too small to use all $b$~bits, it is padded with zeros.

\begin{figure}
\begin{minipage}[t]{0.5\linewidth}\small
\begin{alltt}
\textbf{struct} Fields4\_8  \{
    unsigned    Int1: 4;
    unsigned    Int2: 4;
    unsigned    Int3: 4;
    unsigned    Int4: 4;
    unsigned    Int5: 4;
    unsigned    Int6: 4;
    unsigned    Int7: 4;
    unsigned    Int8: 4;
\};
\end{alltt}
\end{minipage}
\begin{minipage}[t]{0.5\linewidth}\small
\begin{alltt}
\textbf{struct} Fields5\_8 \{
  unsigned    Int1: 5;
  unsigned    Int2: 5;
  unsigned    Int3: 5;
  unsigned    Int4: 5;
  unsigned    Int5: 5;
  unsigned    Int6: 5;
  unsigned    Int7: 5;
  unsigned    Int8: 5;
\};
\end{alltt}
\end{minipage}
\caption{\label{fig:bitfields}Eight bit-packed integers represented as two structures in C/C++. 
Integers in the left panel use 4-bit fields, while integers in the right panel use 5-bit fields.}
\end{figure}

Languages like C and C++ support the concept of bit packing through bit fields. 
An example of two C/C++ structures with bit fields is given in Fig.~\ref{fig:bitfields}. 
Each structure in this example stores 8 small integers.
 The structure \texttt{Fields4\_8} uses 4~bits per integer ($b=4$),
while the structure \texttt{Fields5\_8} uses 5~bits per integer ($b=5$). 

\begin{figure}\centering
\includegraphics[width=\textwidth]{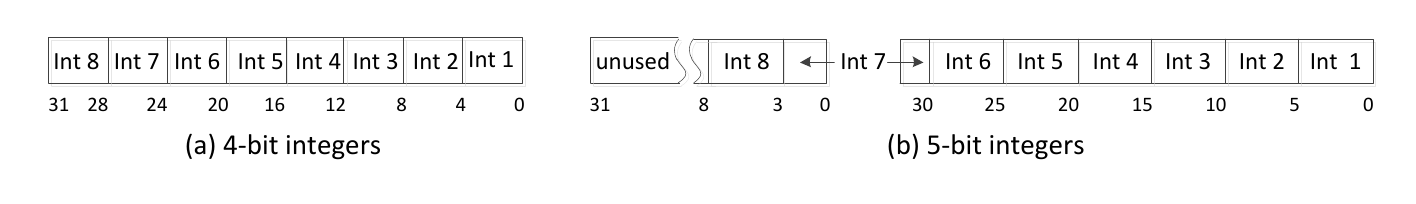}
\caption{\label{fig:bitpack-layout-ex1} Example of two bit-packed representations of 8 small integers.
For convenience, we indicate a starting bit number for each field (numeration begins from zero). 
Integers in the left panel use 4-bit each and, consequently, they fit into a single 32-bit word.
Integers in the right panel use 5-bit each. The complete representation uses two 32-bit words:
24-bits are unoccupied. 
}
\end{figure}

Assuming that bit fields in these structures are stored compactly, i.e., without gaps,
and the order of the bit fields is preserved,
the 8 integers are stored in the memory as shown in Fig.~\ref{fig:bitpack-layout-ex1}.
If any bits remain unused, their values can be arbitrary.
All small integers on the left panel in Fig.~\ref{fig:bitpack-layout-ex1} fit into a single 32-bit word.
However, the integers on the right panel require two 32-bit words with 24~bits remaining unused 
(these bits can be arbitrary). The field of the 7\addth integer crosses the 32-bit word boundary:
the first 2~bits use bits 30--31 of the first words, while
the remaining 3~bits occupy bits 0--2 of the second word (bits are enumerated starting from zero). 

Unfortunately, language implementers are not required to ensure that the data is fully packed. For example,  the C language specification states that \emph{whether a bit-field that does not fit is put into the next unit or overlaps adjacent units is implementation-defined}~\cite{Jones:2003:NCS:995693}.
Most importantly, they do not have to provide packing and unpacking routines that are optimally fast. 
Hence, we implemented bit packing and unpacking using our own procedures as proposed by Zukowski et al.~\cite{1617427}.
In Fig.~\ref{fig:bitunpack}, we give C/C++ implementations of such procedures
assuming that fields are laid out as depicted 
in Fig.~\ref{fig:bitpack-layout-ex1}.
The packing procedures can be implemented similarly and
we omit them for simplicity of exposition.

In some cases, we use bit packing even though some integers are larger than $2^b - 1$ (see \S~\ref{sec:patchedcoding}). In effect, we want to  pack only the first $b$~bits of each integer, which   can be implemented by applying a bit-wise logical \texttt{and} operation with the mask $2^b-1$ on each integer. These extra steps slow down the bit packing (see \S~\ref{sec:expbitpacking}).

The procedure \texttt{unpack4\_8} decodes eight 4-bit integers. 
Because these integers are tightly packed, they occupy exactly one 32-bit word.
Given that this word is already loaded in a register, 
each integer can be extracted using at most four simple operations (shift, mask, store, and 
pointer increment). Unpacking is efficient because it does not involve any branching. 

The procedure \texttt{unpack5\_8} decodes eight 5-bit integers.
This case is more complicated, 
because the packed representation uses two words:
the field for the 7\addth integer crosses word boundaries.
The first two (lower order) bits of this integer are stored in the first word,
while the remaining three (higher order) bits are stored in the second word. 
Decoding does not involve any branches 
and most integers are extracted using four simple operations.

The procedures \texttt{unpack4\_8} and \texttt{unpack5\_8} are merely examples. 
Separate procedures are required for each
bit width (not just 4 and 5).
\definecolor{dred}{RGB}{200,0,0}

\begin{figure}
\begin{minipage}[t]{0.5\linewidth}\small
\begin{alltt}
\textbf{void} unpack4_8(\textbf{const} uint32\_t* in,
          uint32_t* out) \{
  *out++ = ((*in))       & \textcolor{dred}{15};
  *out++ = ((*in) >> \textcolor{dred}{4})  & \textcolor{dred}{15};
  *out++ = ((*in) >> \textcolor{dred}{8})  & \textcolor{dred}{15};
  *out++ = ((*in) >> \textcolor{dred}{12}) & \textcolor{dred}{15};
  *out++ = ((*in) >> \textcolor{dred}{16}) & \textcolor{dred}{15};
  *out++ = ((*in) >> \textcolor{dred}{20}) & \textcolor{dred}{15};
  *out++ = ((*in) >> \textcolor{dred}{24}) & \textcolor{dred}{15};
  *out =   ((*in) >> \textcolor{dred}{28});
\}
\end{alltt}
\end{minipage}
\begin{minipage}[t]{0.5\linewidth}\small
\begin{alltt}
\textbf{void} unpack5_8(\textbf{const} uint32_t* in,
          uint32_t* out) \{
  *out++ = ((*in))       & \textcolor{dred}{31};
  *out++ = ((*in) >> \textcolor{dred}{5} ) & \textcolor{dred}{31};
  *out++ = ((*in) >> \textcolor{dred}{10}) & \textcolor{dred}{31};
  *out++ = ((*in) >> \textcolor{dred}{15}) & \textcolor{dred}{31};
  *out++ = ((*in) >> \textcolor{dred}{20}) & \textcolor{dred}{31};
  *out++ = ((*in) >> \textcolor{dred}{25}) & \textcolor{dred}{31};
  *out   = ((*in) >> \textcolor{dred}{30});
  ++in;
  *out++ |= ((*in) & \textcolor{dred}{7}) << \textcolor{dred}{2};
  *out =    ((*in) >> \textcolor{dred}{3}) & \textcolor{dred}{31};
\}
\end{alltt}
\end{minipage}
\caption{\label{fig:bitunpack}Two procedures to unpack eight bit-packed integers. 
The procedure \texttt{unpack4\_8} works for $b=4$ while
procedure \texttt{unpack5\_8} works for $b=5$. 
In both cases we assume that (1) integers are packed tightly, i.e., without gaps,
(2) packed representations use whole 32-bit words: values of unused bits are undefined.}
\end{figure}

Decoding routines \texttt{unpack4\_8} and \texttt{unpack5\_8} operate on \emph{scalar} 32-bit values.
An effective way  
to improve performance of these routines 
involves
 \emph{vectorization}~\cite{Lemke:2010:SUQ:1881923.1881936,Willhalm:2009:SUF:1687627.1687671}. 
Consider listings in Fig.~\ref{fig:bitunpack} and 
assume that \texttt{in} and \texttt{out} are pointers to \mbox{$m$-element} vectors instead of scalars.
Further, assume that scalar operators (shifts, assignments, and bit-wise logical operations)
are vectorized. 
For example, a bit-wise shift is applied to all $m$ vector elements at the same time.
Then, a single call to \texttt{unpack5\_8} or \texttt{unpack4\_8} decodes $m\times 8$ rather than just eight integers.

\begin{figure}\centering
\includegraphics[width=1.00\textwidth]{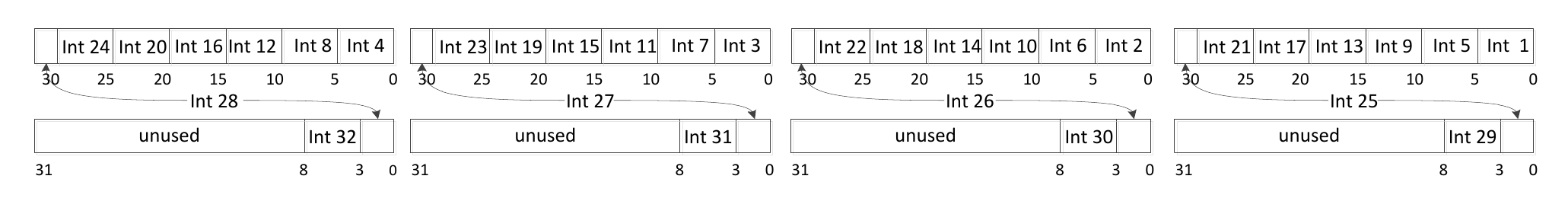}
\caption{\label{fig:bitpack-layout-ex2} 
Example of a vectorized bit-packed representations of 32 small integers.
For convenience, we show a starting bit number for each field (numeration begins from zero).
Integers use 5-bit each. 
Words in the second row follow (i.e., have larger addresses) words of the first row.
Curved lines with arrows indicate that integers \mbox{25--28} are each split between two words.
}
\end{figure}

Recent x86 processors have SIMD instructions that operate on vectors of four~32-bit integers ($m=4$)~\cite{Zhou:2002:IDO:564691.564709,Inoue:2012:HSA:2232971.2232976,SPE:SPE1109}. 
We can use these instructions to achieve a better decoding speed.
A sample vectorized data layout for $b=5$ is given in Fig.~\ref{fig:bitpack-layout-ex2}. 
Integers are divided among series of four 32-bit words in a round-robin fashion.
When a series of four words overflows, the data \emph{spills} over to the next series of 32-bit integers.
In this example, the first 24 integers are stored in the first four words 
(the first row in Fig.~\ref{fig:bitpack-layout-ex2}), 
integers 25--28 are each split between different words, and the remaining integers 29--32 are stored in the second
series of words (the second row of the Fig.~\ref{fig:bitpack-layout-ex2}). 

These data can be processed using a vectorized version of the procedure \texttt{unpack5\_8},
which is obtained from \texttt{unpack5\_8} by  replacing scalar
operations with respective SIMD instructions. 
With Microsoft, Intel or GNU GCC compilers, we can almost mechanically go from the scalar procedure to the vectorized one by replacing each C operator with the equivalent SSE2 intrinsic function: \begin{itemize}
\item the bitwise logical and (\&) becomes 
\texttt{\_mm\_and\_si128},
\item  the right shift ($>>$) becomes \texttt{\_mm\_srli\_epi32}, 
\item and the left shift ($<<$) becomes \texttt{\_mm\_slli\_epi32}. 
\end{itemize}
Indeed, compare procedure \texttt{unpack5\_8} from Fig.~5
with procedure \texttt{SIMDunpack5\_8} from Fig.~7.
The intrinsic functions serve the same purpose as the C operators except
that they work on vectors of 4~integers instead of single integers: e.g., the function 
\texttt{\_mm\_srli\_epi32} shifts 4~integers at once.
The functions \texttt{\_mm\_load\_si128} and \texttt{\_mm\_store\_si128} load a register from memory and write the content of a register to memory respectively; the function 
\texttt{\_mm\_set1\_epi32} creates a vector of 4~integers initialized with a single integer (e.g., 31 becomes 31,31,31,31).

In the beginning of the vectorized procedure  
the pointer \texttt{in} points to the first 128-bit chunk of data displayed in row one of the Fig.~\ref{fig:bitpack-layout-ex2}. The first shift-and-mask operation extracts 4~small integers at once. 
Then, these integers are written to the target buffer using a \emph{single} 128-bit SIMD store operation.
The shift-and-mask is repeated until we extract the first 24 numbers and the first two bits of 
the integers 25--28. At this point the unpack procedure increases the pointer \texttt{in} and
loads the next 128-bit chunk into a register.
Using an additional mask operation, it extracts the remaining 3~bits of 
integers 25--28. These bits are combined with already obtained first 2~bits (for each
of the integers 25--28).  Finally, we store integers 25--28 and finish processing
the second 128-bit chunk by extracting numbers 29--32.

\begin{figure}
\small
\begin{alltt}
const static __m128i m7  =  \emph{_mm_set1_epi32}(\textcolor{dred}{7}U);
const static __m128i m31 =  \emph{_mm_set1_epi32}(\textcolor{dred}{31}U); 

\textbf{void} SIMDunpack5_8(\textbf{const} __m128i* in,  __m128i* out) \{
  __m128i i = \emph{_mm_load_si128}(in);
  \emph{_mm_store_si128}(out++, \emph{_mm_and_si128}( i , m31)); 
  \emph{_mm_store_si128}(out++, \emph{_mm_and_si128}(  \emph{_mm_srli_epi32}(i,\textcolor{dred}{5})  , m31));
  \emph{_mm_store_si128}(out++, \emph{_mm_and_si128}(  \emph{_mm_srli_epi32}(i,\textcolor{dred}{10}) , m31)); 
  \emph{_mm_store_si128}(out++, \emph{_mm_and_si128}(  \emph{_mm_srli_epi32}(i,\textcolor{dred}{15}) , m31)); 
  \emph{_mm_store_si128}(out++, \emph{_mm_and_si128}(  \emph{_mm_srli_epi32}(i,\textcolor{dred}{20}) , m31)); 
  \emph{_mm_store_si128}(out++, \emph{_mm_and_si128}(  \emph{_mm_srli_epi32}(i,\textcolor{dred}{25}) , m31)); 
  __m128i  o = \emph{_mm_srli_epi32}(i,\textcolor{dred}{30}); 
  i = \emph{_mm_load_si128}(++in); 
  o = \emph{_mm_or_si128}(o, \emph{_mm_slli_epi32}(\emph{_mm_and_si128}(i, m7), \textcolor{dred}{2}));
  \emph{_mm_store_si128}(out++, o); 
  \emph{_mm_store_si128}(out++, \emph{_mm_and_si128}(  \emph{_mm_srli_epi32}(i,\textcolor{dred}{3})  , m31)); 
\}
\end{alltt}
\caption{\label{fig:simidbitunpack}Equivalent
to the \texttt{unpack5\_8} procedure from Fig.~\ref{fig:simidbitunpack}
using SSE~intrinsic functions as illustrated by Fig.~\ref{fig:bitpack-layout-ex2}.
}
\end{figure}

\begin{figure}\centering
\includegraphics[width=\textwidth]{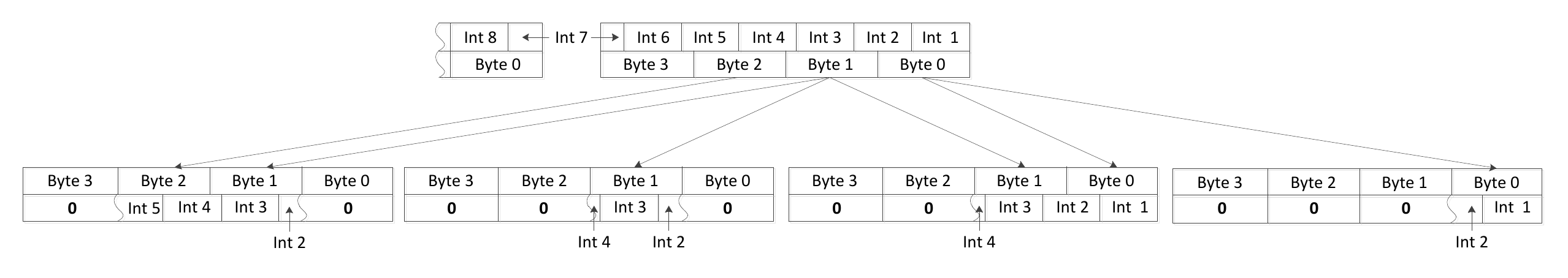}
\caption{\label{fig:interl-simd} One step of simultaneous decoding of four 5-bit integers
that are stored in a horizontal layout (as opposed to the vertical data layout of Fig.~\ref{fig:bitpack-layout-ex2}).
These integers are copied to four 32-bit words using the shuffle operation \texttt{pshufb}.
The locations in source and target buffers are indicated by arrows. 
Curvy lines are used to denote integers that cross byte boundaries in the source buffer. 
Hence, they are copied only partially. 
The boldface zero values represent the bytes zeroed by the shuffle instruction.
Note that some source bytes are copied to multiple locations. 
}
\end{figure}

Our vectorized data layout is interleaved. That is, the first four integers (Int~1, Int~2, Int~3, and Int~4 in Fig.~\ref{fig:bitpack-layout-ex2}) are packed into 4~different 32-bit words. The first integer is immediately adjacent to the fifth integer (Int~5). Schlegel et al.~\cite{Schlegel:2010:FIC:1869389.1869394} called this model \emph{vertical}. Instead we could ensure that the integers are packed sequentially (e.g. Int~1, Int~2, and Int~3 could be stored in the same 32-bit word).  Schlegel et al.\ called this alternative model \emph{horizontal} and it is used by 
Willhalm et al.~\cite{Willhalm:2009:SUF:1687627.1687671}. 
In their scheme, decoding relies on the SSSE3 shuffle operation \texttt{pshufb} (like varint-G8IU).
After we determine the bit width $b$ of integers in the block,
one decoding step typically includes the following operations:
\begin{enumerate}
\item Loading data into the source 16-byte buffer (this step may require a 16-byte alignment).
\item Distributing 3--4 integers stored in the source buffer among
four 32-bit words of the target buffer.
This step, which requires loading a shuffle mask, is illustrated by  Fig.~\ref{fig:interl-simd}
(for 5-bit integers).
Note that unlike varint-G8IU, the integers in the source buffer are not necessarily aligned by byte boundaries
 (unless $b$ is 8, 16, or 32).
Hence, after the shuffle operation,
(1)~the integers copied the target buffer may not be aligned on boundaries of 32-bit words, and (2)~\mbox{32-bit} words may contain some extra bits that do not belong to the integers of interest.
\item Aligning integers on bit boundaries, which may require shifting  several integers to the right.
Because the x86 platform currently 
lacks a SIMD shift that has four different shift amounts, this step is simulated
via  two operations: a SIMD multiplication by four different integers using the SSE4.1~instruction \texttt{pmulld}  and a subsequent vectorized right shift. 
\item Zeroing bits that do not belong to the integers of interest. This requires a mask operation.
\item Storing the target buffer.
\end{enumerate}
Overall, Willhalm et al.~\cite{Willhalm:2009:SUF:1687627.1687671} require SSE4.1 for their horizontal bit packing whereas efficient bit packing using a vertical layout  only requires SSE2.

We compare experimentally vertical and horizontal bit packing in \S~\ref{sec:expbitpacking}.

\section{Novel schemes: SIMD-FastPFOR, FastPFOR and SimplePFOR}
\label{sec:fastpfor}

Patched schemes compress arrays broken down into 
pages (e.g., thousands or millions of integers).
Pages themselves may be broken down into small blocks (e.g., 128~integers). While the original patched coding scheme (PFOR) stores exceptions on a per page basis, newer alternatives such as NewPFD and OptPFD store exceptions
on a per block basis (see Table~\ref{table:overviewpatched}). Also, PFOR picks a single bit width for an entire page, whereas NewPFD and OptPFD   may choose a separate bit width for each block. 

The net result is that NewPFD compresses better than PFOR, but PFOR is faster than NewPFD\@. We would prefer a  scheme that compresses as well as  NewPFD but with the speed of PFOR\@. For this purpose, we propose two new schemes: FastPFOR and SimplePFOR\@. Instead of compressing the exceptions on a per block basis like NewPFD and OptPFD, FastPFOR and SimplePFOR  store the exceptions  on a per page basis, which is similar to PFOR\@. However, like NewPFD and OptPFD, they pick a new bit width for each block.

To explain FastPFOR and SimplePFOR, we consider an example. For simplicity, we only use 16~integers (to be encoded). In the binary notation these numbers are: 
\begin{eqnarray*}
10, 10, 1, 10, 100110, 10, 1, 11, 10, 100000, 10, 110100, 10, 11, 11, 1
\end{eqnarray*}
The maximal number of bits used by an integer  is 6 (e.g., because of 100000). So we can store the data using 6~bits per value plus some overhead. However, we might be able to do better by allowing exceptions in the spirit of patched coding. Assume that we store the location of any exception using a byte (8~bits): in our  implementation, we use blocks of 128~integers so that this is not a wasteful choice.

We want to pick $b\leq 6$, the actual number of bits
we use. That is, we store the lowest $b$~bits of each value.
If a value uses $6$~bits, then we somehow need to store the extra $6-b$~bits as an exception.
We propose to use the difference (i.e., $6-b$) between the maximal bit width and the number of 
bits allocated per  truncated integer  to estimate the cost of storing an exception. 
This is a heuristic since we use slightly more in practice (to compress the $6-b$~highest bits of exception values). 
Because we store exception locations using 8~bits,
 we estimate the cost of storing each exception as $8+(6-b)=14-b$~bits. 
 We want to choose $b$ so that $b\times 16 + (14-b) \times c$ is minimized where $c$ is the number of exceptions corresponding to the value $b$. (In our  software,
we store blocks of 128~integers so that the formula would be $b\times 128 + (14-b) \times c$.) 

We still need to compute the number of exceptions $c$ as a function of the bit width $b$ in a given 
block of integers.
For this purpose,  we  build a histogram 
that tells us how many  integers have a given bit width. In software, this can be implemented as an array of 33~integers: one integer for each possible bit width from 0 to 32.   Creating the histogram requires the computation of the integer logarithm ($\lceil \log_2 (x+1) \rceil$) of every single integer to be coded. 
From this histogram, we can quickly determine the value $b$  that minimizes the expected storage simply by trying every possible value of $b$. 
 Looking at our data, we have 3~integers using 1~bit, 10~integers using 2~bits, and 3~integers using 6~bits.
 So, if we set $b=1$, we get $c=13$~exceptions; for $b=2$, we get $c=3$;  and for $b=6$, $c=0$. The corresponding costs ($b\times 16 + (14-b) \times c$) are 185, 68, and 96. So, in this case, we choose $b=2$. We therefore have 3~exceptions (100110, 100000, 110100).

A compressed page begins with a 32-bit integer. Initially, this 32-bit integer is left uninitialized: we come back to it later. 
Next,  we first store the values themselves, with the restriction that we use only $b$~lowest bits of each value. In our example, the data corresponding to the block is
\begin{eqnarray*}
10, 10, 1, 10, {\color{red}10}, 10, 1, 11, 10, {\color{red}00}, 10, {\color{red}00}, 10, 11, 11, 1.
\end{eqnarray*}
These truncated values are stored continuously, one block after the other (see Fig.~\ref{fig:fastpfor}). Different blocks may use different values of $b$, but because $128 \times b$ is always divisible by $32$, the truncated values for a given block can be stored at a memory address that is 32-bit aligned.

During the encoding of a compressed page, we write to a temporary byte array.
The byte array contains different types of information.
For each block, we store the number of bits allocated for each truncated integer (i.e., $b$) 
and  the maximum number of bits any actual, i.e., non-truncated, value may use. 
If the maximal bit width is greater than the number of allocated bits $b$,
 we store a counter $c$ indicating the number of exceptions. 
 We also store the $c$~exception locations within the block as integers in $[0,127]$.
In contrast with schemes such as NewPFD or OptPFD, 
we do not attempt to compress these numbers and simply store them using one byte each.
Each value is already represented concisely using only one byte as opposed to using a 32-bit integer or worse.

When all integers of a page have been processed and bit-packed,
the temporary byte array 
is stored right after the truncated integers, 
preceded with a 32-bit counter indicating its size. We pad the byte array with zeros so that the number of bytes is divisible by 4 (allowing a 32-bit memory alignment). 
Then, we go back to the beginning of the compressed page, where we had left an uninitialized 32-bit integer, and we write there the offset of the byte array within the compressed page. This ensures that during decoding we can locate the  byte array immediately.
The initial 32-bit integer and the 32-bit counter preceding the byte array add a fixed overhead of 8~bytes per page: it is typically negligible because it is shared by many blocks, often spanning thousands of integers. 

In our example, we write 16 truncated integers using $b=2$ bits each, for the total of 4 bytes~(32~bits).
In the byte array, we store:  
\begin{itemize}
\item the number of bits  ($b=2$) allocated per  truncated integer 
using one byte; 
\item the maximal bit width (6) using a byte; 
\item the number of exceptions $c=3$ (again using one byte); 
\item locations of the exceptions (4,9,11) using one byte each.
\end{itemize}
Thus, for this block alone, we use 3+4+3=10~bytes (80~bits). 

 Finally, we must store the highest $(6-b)=4$~bits of each exception:  1001,  1000, 1101. They are stored right after the byte array. 
Because the offset of the byte array within the page is stored at the beginning of the page, and because the byte array is stored with a header indicating its length, we can locate the exceptions quickly during decoding.  The exceptions are stored on a per page basis in compressed form. This is in contrast to schemes such as OptPFD and NewPFD where exceptions are stored on a per-block basis, interleaved with the truncated values.  
 
 SimplePFOR and FastPFOR differ in how they compress high bits of the exception values:

\begin{itemize}
\item 
In the SimplePFOR scheme, we collect all these values (e.g., such as 1001,  1000, 1101) in one 32-bit array and we compress them using Simple-8b. We apply Simple-8b only once per compressed page.
\item In the FastPFOR scheme, we store exceptions in one of 32~arrays, one for each possible bit width (from 1 to 32). When encoding a block, the difference between the maximal bit width and  $b$ determines in which array the exceptions are stored. Each of the 32~arrays is then bit packed using the corresponding bit width. Arrays are padded so that their length is a multiple of 32~integers. 

In our example, the 3 values corresponding to the high bits of exceptions (1001,  1000, 1101) would be stored in the fourth array and bit-packed using 4~bits per value.

In practice, we store the 32~arrays as follows. We start with a 
32-bit bitset:
 each bit of the  bitset corresponds to one array.
 The bit is set to true if the array is not empty and to false otherwise. Then all non-empty bit-packed arrays are stored in sequence. 
Each bit-packed  array is preceded by a 32-bit counter indicating its length.
\end{itemize}
In all other aspects, SimplePFOR and FastPFOR are identical.

These schemes provide effective compression even though they were designed for speed. Indeed,
suppose we could compress the highest bits of 3~exceptions of our example (1001,  1000, 1101) using only 4~bits each. For this block alone, we use 32~bits for the truncated data, 48~bits in the byte array plus 12~bits for the values of the exceptions. 
The total would be 92~bits to store the 16~original integers, or 5.75~bits/int. This compares favorably to maximal bit width of these integers (6). In our  implementation, we use blocks of 128~integers instead of only 16~integers so that good compression is more likely.

During decoding, the exceptions are first decoded in bulk. To ensure that we do not overwhelm the CPU cache, we process the data in pages of $2^{16}$~integers. We then unpack the integers and apply patching on a per block basis. 
The exceptions locations
do not need any particular decoding: they are read byte by byte.

Though SimplePFOR and FastPFOR are similar in design to NewPFD and OptPFD, we find that they offer better coding and decoding speed. In our tests (see \S~\ref{sec:expe}), 
FastPFOR and SimplePFOR encode integers about twice as fast as NewPFD. 
It is an indication that compressing  exceptions in bulk is faster.

\begin{figure}
\small
\begin{tabular}{lc}\hline
\rule{0pt}{0.5cm} 
Data to be compressed: & \ldots 10, 10, 1, 10, 100110, 10, 1, 11, 10, 100000, 10, 110100, 10, 11, 11, 1\ldots\\[0.3cm]\hline
\rule{0pt}{0.5cm} 
Truncated data: \\($16\times 2 = 32$ bits) & \ldots 10, 10, 01, 10, {\color{red}10}, 10, 01, 11, 10, {\color{red}00}, 10, {\color{red}00}, 10, 11, 11, 01 \ldots \\
\rule{0pt}{0.5cm} 
Byte array:\\ ($6\times 8 = 48$~bits) & \ldots 2, 6, 3, {\color{blue}4, 9, 11} \ldots \\
\rule{0pt}{0.5cm} 
Exception data:\\ (to be compressed)& \ldots 1001,  1000, 1101 \ldots \\[0.3cm]\hline
\end{tabular}
\caption{\label{fig:fastpfor}Layout of a compressed page for SimplePFOR and FastPFOR schemes with our running example. We only give numbers for a block of 16~integers: a page contains hundreds of blocks. The beginning of each page contains the truncated data of each block.  The truncated data is then followed by a byte array containing metadata (e.g., exception locations). 
At the end of the page, we store the exceptions in compressed form.}
\end{figure}

We also designed a new scheme, SIMD-FastPFOR: it is identical to FastPFOR except that 
it packs relies on vectorized bit packing for
the truncated integers and the high bits of the exception values. The compression ratio is slightly diminished for two reasons: 
\begin{itemize}
\item The 32~exception arrays are padded so that their length is a multiple of 128~integers, instead of 32~integers.
\item We insert some padding prior to storing bit packing data so that alignment on 128-bit boundaries is preserved.  
\end{itemize}
This padding adds an overhead of about 0.3--0.4~bits per integer (see Table~\ref{table:aggrealistics}).

\section{Experiments}
\label{sec:expe}

The goal of our experiments is to  evaluate the best known 
integer encoding methods.
The  first series of our test in \S~\ref{sec:synth} is based on synthetic data
sets first presented by Anh and Moffat~\cite{Anh:2010:ICU:1712666.1712668}: ClusterData and Uniform. They have the benefit that they can be
 quickly implemented, thus helping reproducibility.
We then confirm our results in \S~\ref{sec:realistic} using large realistic data sets based
on TREC collections ClueWeb09 and GOV2.

\subsection{Hardware}

We carried out our experiments on a Linux server equipped with Intel Core~i7~2600 (3.40\,GHz, 8192\,KB of L3~CPU cache) and 16\,GB of RAM\@. 
The DDR3-1333 RAM with dual channel
has a transfer rate of  $\approx$20,000\,MB/s or  $\approx$5300\,mis. 
According to our tests, it can copy arrays at a rate of 2270\,mis with the C function \texttt{memcpy}.

\subsection{Software}
\label{sec:soft}
We implemented our algorithms in C++ using GNU~GCC~4.7. We use the optimization flag \texttt{-O3}.  Because the varint-G8IU scheme requires  SSSE3~instructions, we had to add the flag \texttt{-mssse3}. When compiling our implementation of Willhalm et al.~\cite{Willhalm:2009:SUF:1687627.1687671} bit unpacking, we had to use the flag \texttt{-msse4.1} since it requires SSE4~instructions. Our complete source code is available online.\footnote{\url{https://github.com/lemire/FastPFOR}}

Following Stepanov et al.~\cite{Stepanov:2011:SDP:2063576.2063627}, we compute speed based on the wall-clock in-memory processing. 
Wall-clock times include the time necessary for \emph{differential coding and decoding}. 
During our tests, we do not retrieve or store data on disk: it is impossible to decode billions of integers per second when they are kept on disk.

Arrays containing more than $2^{16}$~integers (256\,KB) are broken down into smaller chunks. 
Each chunk is decoded into two passes. 
In the first pass, we decompress deltas and store each delta value using a 32-bit word.
In the second pass, we carry out an in-place computation of prefix sums.
As noted in \S~\ref{sec:fast-delta}, this approach greatly improves data locality and leads
to 
a significant
 improvement in decoding speed for the fastest schemes.

 Our implementation
of VSEncoding,  NewPFD, and OptPFD is based on software published
by Silvestri and Venturini~\cite{Silvestri:2010:VEC:1871437.1871592}.
They report that their implementation of OptPFD was validated against
an implementation provided by the original authors~\cite{yan2009inverted}.
We implemented varint-G8IU from Stepanov et al.~\cite{Stepanov:2011:SDP:2063576.2063627} 
as well as Simple-8b from Anh and Moffat~\cite{Anh:2010:ICU:1712666.1712668}.
To minimize branching, we implemented Simple-8b using a C++ \texttt{switch case} that selects one of 16~functions, that is, one for each selector value. 
Using a function for each selector value instead of a single function is faster because loop unrolling eliminates branching. (Anh and Moffat~\cite{Anh:2010:ICU:1712666.1712668} referred to this optimization as \emph{bulk unpacking}.)
  We also implemented the original PFOR scheme from Zukowski et al.~\cite{1617427} as well as its successor PFOR2008 from Zhang et al.~\cite{1367550}. Zukowski et al.\ made a distinction between PFOR and PFOR-Delta: we effectively use FOR-Delta since we apply PFOR to deltas. 

Reading and writing unaligned data can be as fast as reading and writing aligned data on recent Intel processors---as long as we do not cross a 64-byte cache line. 
Nevertheless,  we still wish to align data on 32-bit boundaries when using regular \emph{binary packing}.
Each block of 32 bit-packed integers should be preceded by a descriptor that stores the bit width ($b$) of integers in the block.
The number of bits used by the block is always divisible by 32.
Hence, to keep blocks aligned on 32-bit boundaries, 
we group the blocks and respective descriptors into meta-blocks each of which contains
4~successive blocks.
A meta-block is preceded by a 32-bit descriptor that combines 4~bit widths $b$ (8~bits per width).
We call this scheme BP32.
We also experimented with versions of binary packing on fewer integers (8~integers and 16~integers).
Because these versions are slower, we omit them from our experiments.

We also implemented a vectorized binary packing over blocks of 128~integers (henceforth \mbox{SIMD-BP128}). 
Similar to regular binary packing, 
we want to keep the blocks aligned on 128-bit boundaries when using vectorized binary packing.
To this end,
we regroup 16~blocks into a meta-block of 2048~integers.
As in BP32, the encoded representation of a meta-block is preceded by a 128-bit descriptor word keeping bit widths (8~bits per width). 

In summary, the format of our  binary packing schemes is as follows:
\begin{itemize}
\item SIMD-BP128 combines 16~blocks of 128~integers whereas BP32 combines 4~blocks of 32~integers.
\item SIMD-BP128 employs (vertical) vectorized bit packing whereas BP32 relies on the regular bit packing as described in \S\ref{sec:fast-bit-unpacking}.
\end{itemize}

Many schemes such as BP32 and SIMD-BP128 require the computation of the integer logarithm during
encoding. 
If done naively, this step can take up most of the running time: the computation of the integer logarithm is slower than a fast operation such as a shift or an addition. 
We found it best
to use  the \emph{bit scan reverse} (\texttt{bsr}) assembly instruction on x86 platforms  (as it provides $ \lceil \log_2 (x + 1)\rceil - 1 $ whenever $x>0$).

For the binary packing schemes,  we must determine the maximum of the integer logarithm of the integers 
($\max_i \lceil \log_2 (x_i + 1) \rceil $)  during encoding.
Instead on computing one integer logarithm per integer, we carry out a bit-wise logical \texttt{or} on all the integers and compute the integer logarithm of the result.
This shortcut is possible due to the equation:
$\max_i \lceil \log_2 (x_i+1) \rceil =\lceil \log_2 \lor_i (x_i+1) \rceil$ where $\lor$ refers to the bit-wise logical \texttt{or}.

Some schemes compress data in blocks of fixed length (e.g.,  128~integers).  We  compress the remainder using Variable~Byte as in Zhang et al.~\cite{1367550}. In our tests, most arrays are  large compared to the block size.
Thus, replacing Variable~Byte by another scheme would make no or little difference.

 Speeds are reported in millions of 32-bit integers per second (mis). Stepanov et al.\ report a speed of 1059\,mis over the TREC GOV2 data set for their best scheme varint-G8IU.\@ We got a similar speed (1300\,mis). 

VSEncoding, FastPFOR, and SimplePFOR use buffers during compression and decompression proportional to the size of the array. VSEncoding uses a persistent buffer of over 256\,KB.\@ We implemented SIMD-FastPFOR, FastPFOR, and SimplePFOR with a persistent buffer of slightly more than 64\,KB.\@  PFOR, PFOR2008, NewPFD, and OptPFD are implemented using persistent buffers proportional to the block size (128~integers in our tests): less than 512\,KB in persistent buffer memory are used for each scheme. Both PFOR and PFOR2008 use pages of $2^{16}$~integers or 256\,KB.\@ During compression, PFOR, PFOR2008, SIMD-FastPFOR, FastPFOR, and SimplePFOR use a buffer to store exceptions. These buffers are limited by the size of the pages and they are released immediately after decoding or encoding an array. 

The implementation of VSEncoding~\cite{Silvestri:2010:VEC:1871437.1871592} uses some SSE2~instructions through assembly during bit unpacking.
Varint-G8IU  makes explicit use of SSSE3 instructions through SIMD intrinsic functions whereas SIMD-FastPFOR and SIMD-BP128 similarly use SSE2~intrinsic functions. 

Though we tested vectorized differential coding with all schemes, we only report results for schemes that
make explicit use of SIMD instructions (SIMD-FastPFOR, SIMD-BP128, and varint-G8IU).
To ensure fast vector processing, we align all initial pointers on 16-byte boundaries.

\subsection{Computing bit packing}
\label{sec:expbitpacking}

We implemented bit packing using hand-tuned functions as originally proposed by 
Zukowski et al.~\cite{1617427}. Given a bit width~$b$, a sequence of $K$~unsigned 32-bit integers are coded to $\lceil K b / 32 \rceil $~integers.
In our tests, we used $K=32$ for the regular version, and $K=128$ for the vectorized version.

\begin{figure}\centering
\subfloat[Optimized but portable C++\label{fig:bitpackingcpp}]{%
\includegraphics[width=0.45\textwidth]{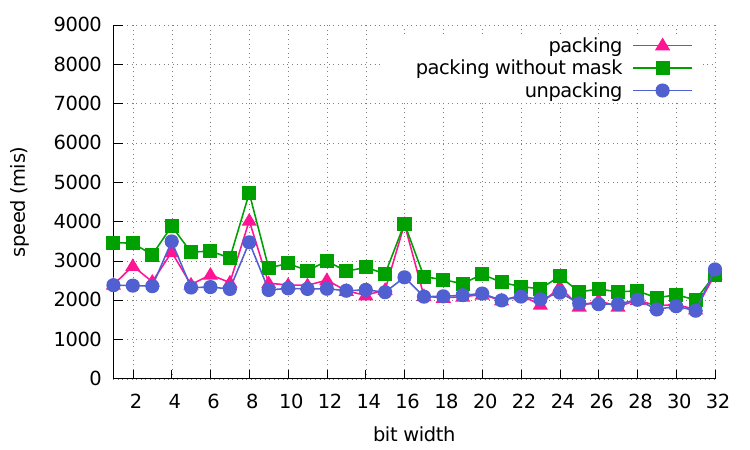}}
\subfloat[Vectorized with  SSE2 instructions\label{fig:bitpackingsse}]{%
\includegraphics[width=0.45\textwidth]{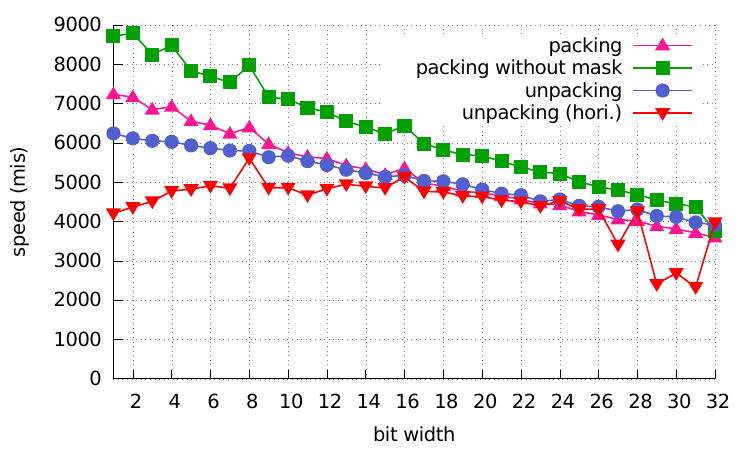}
}

\subfloat[Ratio vectorized/non-vectorized]{%
\includegraphics[width=0.45\textwidth]{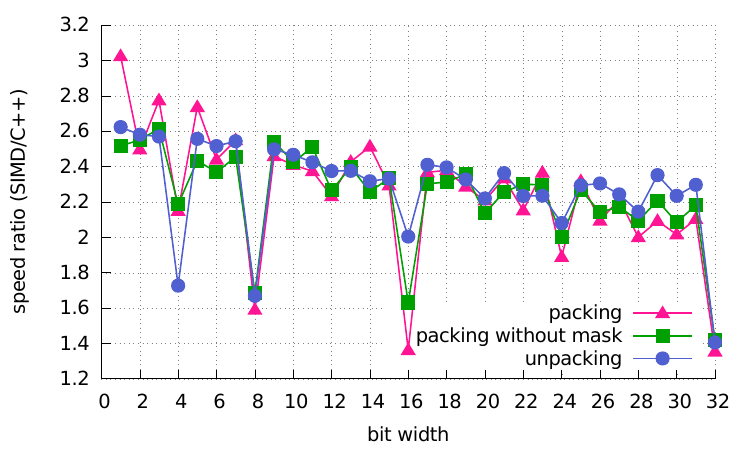}}

\caption{\label{fig:bitpacking}Wall-clock speed in millions of integers per second for bit packing and unpacking. We use small arrays (256\,KB) to minimize cache misses. When packing integers that do not necessarily fit in $b$~bits,
(as required in patching schemes), we must apply a mask which slows down packing by as much as 30\%. 
}
\end{figure}

 Fig.~\ref{fig:bitpacking} illustrates the speed at which we can pack and unpack integers using blocks of 32~integers. 
In some schemes, it is known that all integers are no larger than $2^b - 1$, 
while in patching schemes there are exceptions, i.e., integers larger than or equal to $2^b$.
In the latter case, 
we enforce that integers are smaller than $2^b$ through the application of a mask.
This operation slows down compression.

We can pack and unpack much faster when the number of bits is small because less data needs to be retrieved from RAM\@. Also, we can pack and unpack faster when the bit width is 4, 8, 16, 24 or 32. Packing and unpacking with bit widths of 8 and 16 is especially fast. 
 
 The vectorized version (Fig.~\ref{fig:bitpackingsse}) is roughly twice as fast as the scalar version. We can unpack integers having a bit width of 8 or less at a rate of $\approx$6000\,mis.
However, it carries the implicit constraint that integers must be packed and unpacked in blocks of at least 128~integers.  
 Packing is slightly faster when the bit width is 8 or 16.

In Fig.~\ref{fig:bitpackingsse} only, we report the unpacking speed when using the horizontal data layout  
as described by 
Willhalm et al.~\cite{Willhalm:2009:SUF:1687627.1687671} (see \S~\ref{sec:fast-bit-unpacking}). When the bit widths range from 16 to 26, 
the vertical and horizontal techniques have the same speed.
 For small  ($<8$) or large ($>27$) bit widths, our approach based on a vertical layout is preferable as it is up to 70\% faster. Accordingly, all integer coding schemes are implemented using the vertical layout.
 
 We also experimented with the cases where we pack fewer integers ($K=8$ or $K=16$). However, it is slower and a few bits  remain unused ($\lceil K b / 32 \rceil 32 - K b$). 

\subsection{Synthetic data sets}
\label{sec:synth}

We used  the ClusterData and the Uniform model from Anh and Moffat~\cite{Anh:2010:ICU:1712666.1712668}.
These models generate sets of distinct integers that we keep in sorted order.
In the Uniform model, integers follow a uniform distribution whereas in the ClusterData model, integer values tend  to cluster. That is, we are more likely to get long sequences of similar values. The goal of the ClusterData model is to simulate more realistically data encountered in practice. We expect data obtained  from the ClusterData model to be more  compressible.

We generated data sets of random integers in the range $[0,2^{29})$
with both the ClusterData and the Uniform model.
In the first pass, we generated  $2^{10}$~short arrays containing 
$2^{15}$~integers each. The average difference between successive integers within an array is thus $2^{29-15}=2^{14}$. We expect the compressed data to use at least 14\,bits/int. In the second pass, we generated a single long array of
 $2^{25}$~integers. In this case, the average distance between successive integers is $2^4$: we  expect the compressed data  to use at least 4\,bits/int.

The results are given in Table~\ref{table:synth} 
(schemes with a $^{\star}$ by their name, e.g., SIMD-FastPFOR$^{\star}$, use vectorized differential coding).
Over short arrays, we see little compression as expected. There is also a relatively little difference in compression efficiency between Variable~Byte and 
 a more space-efficient alternative such as FastPFOR\@. 
However, speed differences are large: the decoding speed ranges from 220\,mis for Variable~Byte to 2500\,mis for SIMD-BP128$^{\star}$.

For long arrays, there is a greater difference between the compression efficiencies. The schemes with the best compression ratios are SIMD-FastPFOR, FastPFOR, SimplePFOR, Simple-8b, OptPFD\@. Among those, SIMD-FastPFOR is the clear winner in terms of decoding speed. The good compression ratio of OptPFD comes at a price: it has one of the worst encoding speeds. 
In fact, it is 20--50~times slower than SIMD-FastPFOR during encoding.

Though they differ significantly  in implementation,
FastPFOR, SimplePFOR, and \mbox{SIMD-FastPFOR} have equally good compression ratios.
All three schemes have similar decoding speeds,
but SIMD-FastPFOR decodes integers much faster than FastPFOR and SimplePFOR\@. 

In general, encoding speeds vary significantly, but binary packing schemes are the fastest, especially when they are vectorized. Better implementations could possibly help reduce this gap. 

  The version of
 SIMD-BP128 using vectorized differential coding (written SIMD-BP128$^{\star}$) is always  400\,mis faster during
 decoding than any other alternative. Though it does not always offer the best compression ratio, it always matches the
 compression ratio of Variable~Byte.

 The difference between using vectorized differential coding and regular differential coding could amount to up to 2~bits per integer. Yet, typically, the difference is less than 2~bits. 
 For example, SIMD-BP128$^{\star}$ only uses about one extra bit per integer when compared with \mbox{SIMD-BP128}.  The cost of binary packing is determined by the largest delta in a block: 
increasing the average size of the deltas by a factor of 4 does not necessarily lead to a fourfold increase 
in the expected largest integer (in a block of 128~deltas).

Compared to our novel schemes, performance of varint-G8IU is unimpressive.
However, \mbox{variant-G8IU} is about 60\% faster than Variable~Byte while providing a similar compression efficiency.
It is also faster than Simple-8b, though Simple-8b has a better compression efficiency. 
The version with vectorized differential coding (written varint-G8IU$^{\star}$) has poor compression over the  short arrays compared with the regular version (varint-G8IU). 
Otherwise, on long arrays, varint-G8IU$^{\star}$ is significantly faster (from 1300~mis to 1600~mis) than varint-G8IU while compressing just as well.

There is little difference between PFOR and PFOR2008 except that PFOR offers a significantly faster encoding speed.
Among all the schemes taken from the literature, PFOR and PFOR2008 have the best decoding speed in these
 tests: they use a single bit width for all blocks, determined once at the beginning of the compression. However, they are dominated in all metrics (coding speed, decoding speed and compression ratio) by SIMD-BP128 and SIMD-FastPFOR\@.

For comparison, we tested Google Snappy (version 1.0.5) as a delta compression technique. 
Google Snappy is a freely available library
used internally by Google in its database engines~\cite{click2012}. We believe that it is competitive with other fast generic compression libraries such as zlib or LZO\@.  For short ClusterData arrays, we got a decoding speed of 340\,mis and almost no compression  (29\,bits/int.). For long ClusterData arrays, we got a decoding speed of 200\,mis and 14\,bits/int. Overall, Google Snappy has about half the compression efficiency of SIMD-BP128$^{\star}$  while being an order of magnitude slower. 

\begin{table}
\caption{\label{table:synth}Coding and decoding speed in millions of integers per second over synthetic data sets, together with
number of bits per 32-bit integer. Results are given using  two significant digits. Schemes with a $^{\star}$ by their name use vectorized differential coding.}
\centering

\subfloat[ClusterData: Short arrays]{%
\begin{tabular}{lrrc}
 & coding & decoding  & bits/int \\ \hline 
SIMD-BP128$^{\star}$          & \textbf{1700} & \textbf{2500} & 17 \\
SIMD-FastPFOR$^{\star}$               & 380 & 2000 & 16 \\
SIMD-BP128           & 1000 & 1800 & 16 \\
SIMD-FastPFOR                & 300 & 1400 & \textbf{15} \\
PFOR                        & 350 & 1200 & 18 \\
PFOR2008                    & 280 & 1200 & 18 \\
SimplePFOR                  & 300 & 1100 & \textbf{15} \\
FastPFOR                    & 300 & 1100 & \textbf{15} \\
BP32                        & 790 & 1100 & \textbf{15} \\
NewPFD         & 66 & 1100 & 16 \\
varint-G8IU$^{\star}$                               & 160 & 910 & 23 \\
varint-G8IU                               & 150 & 860 & 18 \\
VSEncoding                               & 10 & 720 & 16 \\
Simple-8b                                 & 260 & 690 & 16 \\
OptPFD         & 5.1 & 660 & 15 \\
Variable Byte                             & 300 & 270 & 17 \\
\end{tabular}
}
\subfloat[Uniform: Short arrays]{%
\begin{tabular}{lrrc}
 & coding & decoding  & bits/int \\ \hline 
    & \textbf{1600} & \textbf{2000} & 18 \\
        & 360 & 1800 & 18 \\
         & 1100 & 1600 & 17 \\
             & 330 & 1400 & \textbf{16} \\
                  & 370 & 1400 & 17 \\
                  & 280 & 1400 & 17 \\
                & 330 & 1200 & \textbf{16} \\
                  & 330 & 1200 & \textbf{16} \\
          & 840 & 1200 & 17 \\
     & 64 & 1300 & 17 \\
                            & 140 & 650 & 25 \\
                          & 170 & 870 & 18 \\
                      & 10 & 690 & 18 \\
                   & 260 & 540 & 18 \\
  & 4.6 & 1100 & 17 \\
                        & 240 & 220 & 19 \\
\end{tabular}
}

\centering
\subfloat[ClusterData: Long arrays]{%
\begin{tabular}{lrrc}
 & coding & decoding  & bits/int \\ \hline 
SIMD-BP128$^{\star}$          & \textbf{1800} & \textbf{2800} & 7.0 \\
SIMD-FastPFOR$^{\star}$               & 440 & 2400 & 6.8 \\
SIMD-BP128           & 1100 & 1900 & 6.0 \\
SIMD-FastPFOR                & 320 & 1600 & 5.4 \\
varint-G8IU$^{\star}$                              & 270 & 1600 & 9.1 \\
PFOR                        & 360 & 1300 & 6.1 \\
PFOR2008                    & 280 & 1300 & 6.1 \\
BP32         & 840 & 1300 & 5.8 \\
FastPFOR                    & 320 & 1200 & 5.4 \\
SimplePFOR                  & 320 & 1200 & \textbf{5.3} \\
varint-G8IU                               & 230 & 1300 & 9.0 \\
NewPFD         & 120 & 970 & 5.5 \\
Simple-8b                                 & 360 & 890 & 5.6 \\
VSEncoding                               & 9.8 & 790 & 6.4 \\
OptPFD         & 17 & 750 & 5.4 \\
Variable Byte                             & 880 & 830 & 8.1 \\
\end{tabular}
}
\subfloat[Uniform: Long arrays]{%
\begin{tabular}{lrrc}
 & coding & decoding  & bits/int \\ \hline 
        & \textbf{1900} & \textbf{2600} & 8.0\\
               & 380 & 2200 & 7.6 \\
          & 1100 & 1800 & 7.0 \\
              & 340 & 1600 & 6.4 \\
                            & 270 & 1600 & 9.0 \\
                   & 360 & 1300 & 7.3 \\
                  & 280 & 1300 & 7.3 \\
             & 810 & 1200 & 6.7 \\
               & 330 & 1200 & 6.3 \\
             & 330 & 1200 & 6.3 \\
                     & 230 & 1300 & 9.0 \\
    & 110 & 1000 & 6.5 \\
                        & 370 & 940 & 6.4 \\
     & 9.9 & 790 & 7.2 \\
   & 15 & 740 & \textbf{6.2} \\
      & 930 & 860 & 8.0\\
\end{tabular}
}
\end{table}

\subsection{Realistic data sets}
\label{sec:realistic}
The posting list of a word is an array of document identifiers where the word occurs. 
For more realistic data sets, we used posting lists obtained from two TREC Web collections. 
Our data sets include only document  identifiers, but not positions of words in documents. 
For our purposes, we do not store the words or the documents themselves, just the posting lists.

The first data set is a posting list collection extracted from the ClueWeb09 (Category B) data set~\cite{clueweb09gap}. 
The second data set is a posting list collection built from the GOV2 data set
by  Silvestri and Venturini~\cite{Silvestri:2010:VEC:1871437.1871592}.
The GOV2 is a crawl of the \texttt{.gov} sites, which contains 25 million HTML, text, and PDF documents
(the latter are converted to text).

This ClueWeb09 collection is a more realistic HTML collection of about 50 million crawled HTML documents, mostly in English. 
It represents postings for one million most frequent words.
Common stop words were excluded and different grammar forms of words were conflated.
Documents were enumerated in the order they appear in source files, i.e., they were not reordered.
Unlike GOV2, the ClueWeb09 crawl is not limited to any specific domain. 
Uncompressed, the posting lists from GOV2 and ClueWeb09 use 20\,GB and 50\,GB
respectively.

\begin{figure}\centering
\subfloat[\label{fig:shann}Shannon entropy of the differences (deltas)]{%
\includegraphics[width=0.45\textwidth]{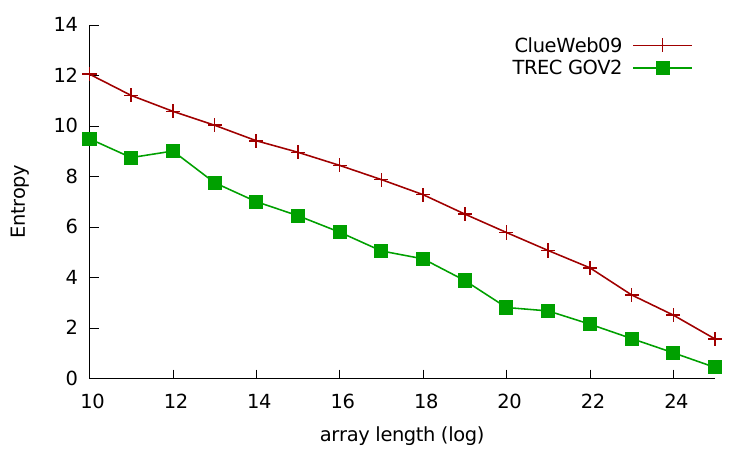}
}
\subfloat[Data distribution\label{fig:datadist}]{%
\includegraphics[width=0.45\textwidth]{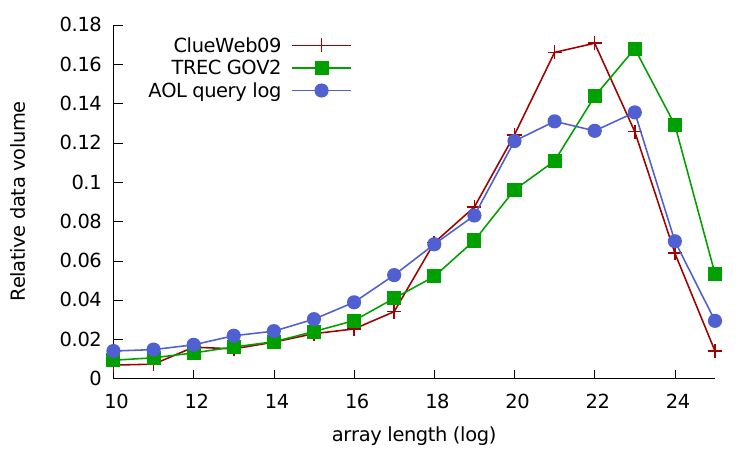}
}
\caption{\label{fig:descr}Description of the posting-list data sets}
\end{figure}

We decomposed these data sets according to the array length,
storing all arrays of lengths  $2^K$ to $2^{K+1}-1$ consecutively. 
We applied differential coding on the arrays (integers $x_1, x_2, x_3, \ldots $ are transformed to $y_1 = x_1, y_2 = x_2-x_1, y_3 = x_3 - x_2, \ldots$)  and computed the Shannon entropy ($-\sum_i p(y_i) \log_2 p(y_i)$) of the result. 
 We estimate the probability $p(y_i)$ of the integer value $y_i$ as the number of occurrences of $y_i$ divided by the number of integers. 
As Fig.~\ref{fig:descr} shows, longer arrays are more compressible. 
There are differences in entropy values between two collections 
(ClueWeb09 has about two extra bits,
see Fig.~\ref{fig:shann}), but these differences are much smaller than those
among different array sizes.
Fig.~\ref{fig:datadist} shows the distribution of array  lengths  and 
 entropy values.

\subsubsection{Results over different array lengths}

\begin{figure}\centering\small
\subfloat[Size: ClueWeb09 (bits/int)\label{fig:sizeclue-bits}]{%
\includegraphics[width=0.45\textwidth]{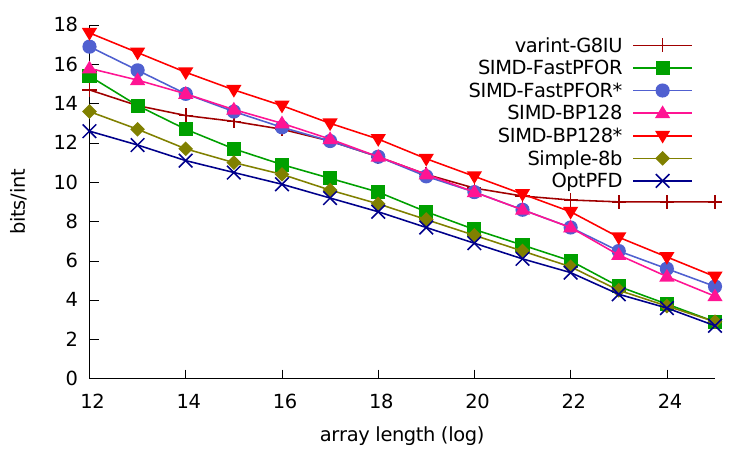}
}
\subfloat[Size: ClueWeb09 (relative to entropy)\label{fig:sizeclue}]{%
\includegraphics[width=0.45\textwidth]{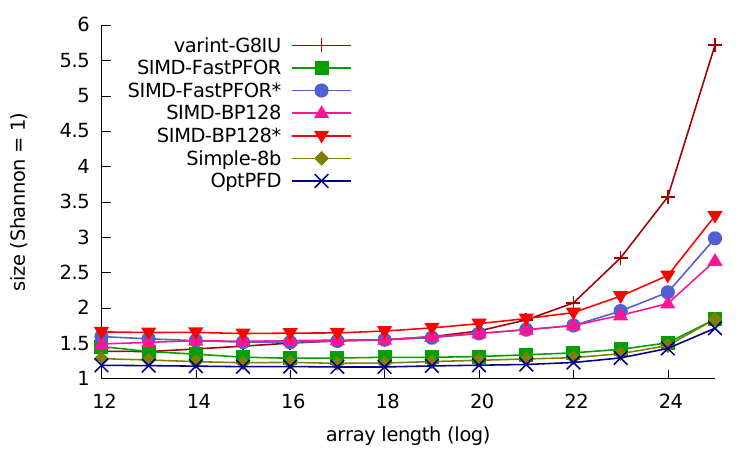}
}
\\
\subfloat[Encoding: ClueWeb09\label{fig:encclue}]{%
\includegraphics[width=0.45\textwidth]{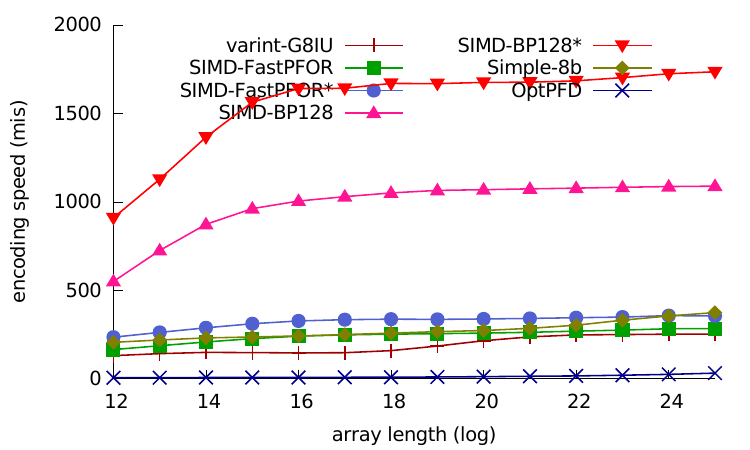}
}
\subfloat[Decoding: ClueWeb09\label{fig:decclue}]{%
\includegraphics[width=0.45\textwidth]{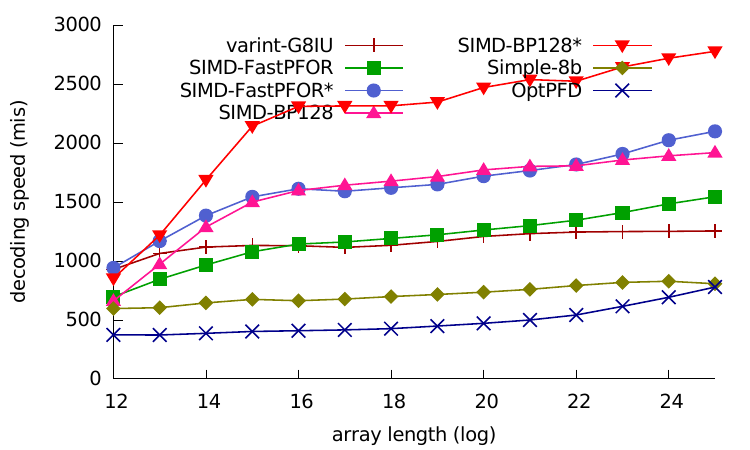}
}\\
\subfloat[Size: GOV2 (bits/int)\label{fig:sizegov2-bits}]{%
\includegraphics[width=0.45\textwidth]{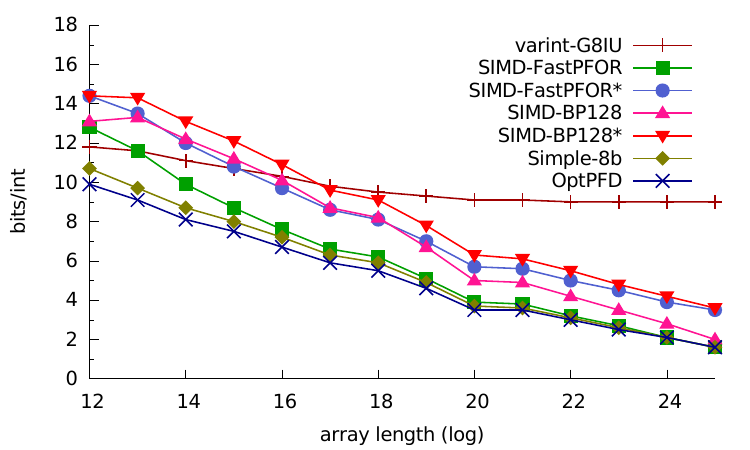}
}
\subfloat[Size: GOV2 (relative to entropy)\label{fig:sizegov2}]{%
\includegraphics[width=0.45\textwidth]{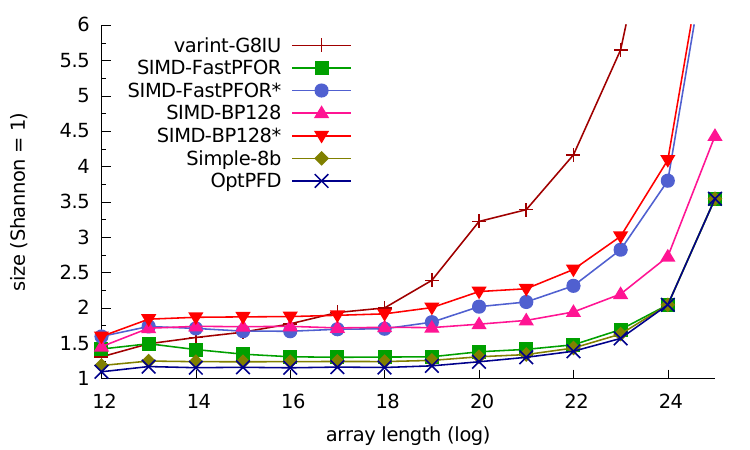}
}
\\
\subfloat[Encoding: GOV2\label{fig:encgov2}]{%
\includegraphics[width=0.45\textwidth]{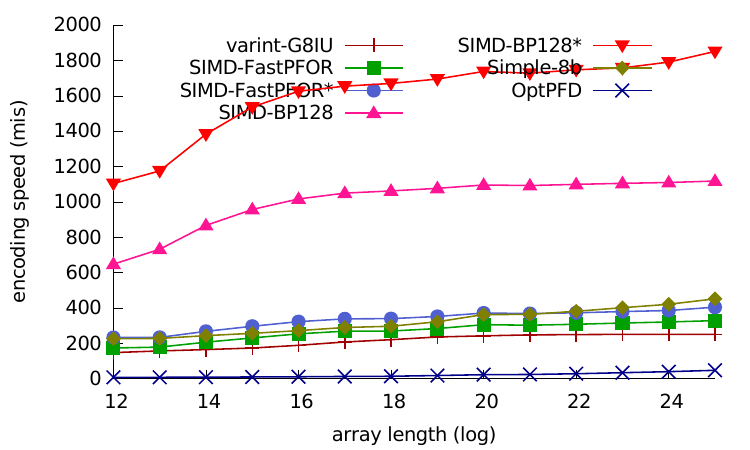}
}
\subfloat[Decoding: GOV2\label{fig:decgov2}]{%
\includegraphics[width=0.45\textwidth]{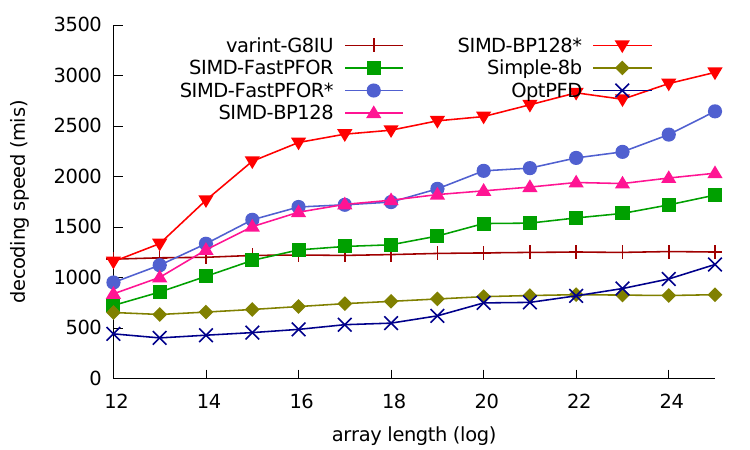}
}
\caption{\label{fig:unaggregated}
Experimental comparison of competitive schemes on ClueWeb09 and GOV2.}
\end{figure}

We present  results per array length for selected schemes in  Fig.~\ref{fig:unaggregated}. 
Longer arrays are more compressible since the deltas, i.e., differences between adjacent elements, are smaller. 

We see in Figs.~\ref{fig:sizeclue} and~\ref{fig:sizegov2} that all schemes compress the deltas within a factor of two of the Shannon entropy for short arrays. For long arrays however, the compression  (compared to the Shannon entropy) becomes worse for all schemes.
Yet many of them manage to remain within a factor of three of the Shannon entropy.

Integer compression schemes are better able to compress close to  the Shannon entropy over ClueWeb09 (see Fig.~\ref{fig:sizeclue}) than over GOV2 (see Fig.~\ref{fig:sizegov2}). For example, SIMD-FastPFOR, Simple-8b, and OptPFD are within a factor of two of Shannon entropy over ClueWeb09 for all array lengths, whereas they all exceed three~times the Shannon entropy over GOV2 for the longest arrays. Similarly, \mbox{varint-G8IU}, SIMD-BP128$^{\star}$, and SIMD-FastPFOR$^{\star}$ remain within a factor of six of the Shannon entropy over ClueWeb09 but they all exceed this factor over GOV2 for long arrays.
In general, it might be easier to compress data close to the entropy when the entropy is larger.

We get poor results with varint-G8IU over the longer (and more compressible)
arrays 
(see Figs.~\ref{fig:sizeclue-bits} and~\ref{fig:sizegov2-bits}). 
We do not find this surprising because variant-G8IU requires at least 9~bits/int. 
In effect, when other schemes such as SIMD-FastPFOR and SIMD-BP128 use less than $\approx 8$~bits/int, they surpass varint-G8IU in both 
compression efficiency and decoding speed. 
However, when the storage exceeds 9~bits/int, Varint-G8IU is one of the fastest methods available  for these data sets. However, we also got  poor results with variant-G8IU on the ClusterData and  Uniform data sets for short (and poorly compressible)  arrays  in \S~\ref{sec:synth}.

We see in Figs.~\ref{fig:encclue} and~\ref{fig:encgov2} that both SIMD-BP128 and SIMD-BP128$^{\star}$ have a significantly better encoding speed, irrespective of the array length. The opposite is true for OptPFD: it is much slower than the alternatives. 

Examining the decoding speed as a function of array length (see Figs.~\ref{fig:encclue} and~\ref{fig:encgov2}), we see that several schemes have a significantly worse decoding speed over short (and poorly compressible) arrays, but the effect is most pronounced for the new schemes we introduced (SIMD-FastPFOR, SIMD-FastPFOR$^{\star}$, \mbox{SIMD-BP128}, and SIMD-BP128$^{\star}$). 
Meanwhile, varint-G8IU and Simple-8b have a decoding speed that is  less sensitive to the array length.

\subsubsection{Aggregated results}
Not all posting lists are equally likely to be retrieved by the search engine. As observed by Stepanov et al.~\cite{Stepanov:2011:SDP:2063576.2063627}, it is desirable to  account for different term distributions in queries. Unfortunately, we do not know of an ideal approach to this problem. 
Nevertheless, 
to model more closely the performance of a major search engine, we used the AOL query log data set as a collection of query statistics~\cite{Brenes:2009:SAA:1523512.1523572,Pass:2006}.
It  consists of about 20~million web queries collected from 650~thousand users over three months:
queries repeating within a single user session were ignored.
When possible (in about 90\% of all cases), we matched the query terms with posting lists in the ClueWeb09 data set and obtained term frequencies (see Fig.~\ref{fig:datadist}). 
This allowed us to estimate
 how often a posting list of length  between $2^K$ to $2^{K+1}-1$ is likely to be retrieved for various values of $K$. This gave us a weight vector that we use to aggregate our results.  
 
 We present aggregated results in Table~\ref{table:aggrealistics}. The results are generally similar to what we obtained with synthetic data. The newly introduced schemes (SIMD-BP128$^{\star}$, SIMD-FastPFOR$^{\star}$, \mbox{SIMD-BP128}, \mbox{SIMD-FastPFOR}) still offer the best decoding speed. We find that  varint-G8IU$^{\star}$  is much faster than varint-G8IU (1500\,mis vs. 1300\,mis over GOV2) even though the compression ratio is the same with a margin of 10\%. PFOR and PFOR2008 offer a better compression than \mbox{varint-G8IU$^{\star}$} but at a reduced speed. However, we find that SIMD-BP128 is preferable in every way to
 varint-G8IU$^{\star}$, varint-G8IU, PFOR, and PFOR2008.
 
\begin{table}
\caption{\label{table:aggrealistics}Experimental results. Coding and decoding speeds
are given in millions of 32-bit integers per second.  Averages are weighted based on AOL query logs. 
}
\centering
\subfloat[ClueWeb09]{%
\begin{tabular}{lrrc}
 & coding & decoding  & bits/int \\ \hline 
SIMD-BP128$^{\star}$          &   \textbf{1600}& \textbf{2300} &11 \\
SIMD-FastPFOR$^{\star}$        &         330 &1700 &9.9\\
SIMD-BP128      &      1000 &1600 &9.5\\
varint-G8IU$^{\star}$&      220 &1400 &12\\
SIMD-FastPFOR                &  250 &1200 &8.1 \\
PFOR2008                   & 260 & 1200 & 10 \\
PFOR                        & 330 & 1200 & 11 \\       
varint-G8IU             & 210 & 1200 & 11 \\             
BP32                  & 760& 1100& 8.3 \\
SimplePFOR          & 240& 980 & 7.7\\
FastPFOR        &  240& 980 &7.8 \\
NewPFD        & 100& 890 &8.3 \\
VSEncoding             &  11 &740 &7.6\\
Simple-8b         &  280 &730 &7.5\\
OptPFD          &14& 500 & \textbf{7.1}\\
Variable Byte     &  570 & 540 & 9.6 \\
\end{tabular}
}
\subfloat[GOV2]{%
\begin{tabular}{lrrc}
 & coding & decoding  & bits/int \\ \hline 
         &    \textbf{1600}   &\textbf{2500}     & 7.6 \\
      &                  350   & 1900         & 7.2 \\
    &    1000   &1700    & 6.3 \\
   &       240 & 1500 & 10 \\
     &   290 & 1400 & 5.3 \\
 &                   250  & 1300  & 7.9 \\
    &                310  & 1300  & 7.9 \\
          &                    230 & 1300 & 9.6 \\
         &       790 & 1200 & 5.5 \\
       &         270 & 1100 & 4.8 \\
         &         270 & 1100 & 4.9 \\
     &  150 & 1000 & 5.2\\
                &              11 & 810 & 5.4\\
                &                340 & 780  &4.8\\
   & 23 & 710 & \textbf{4.5}\\
              &             730 & 680 & 8.7\\

\end{tabular}
}
\end{table}

\begin{figure}\centering
\subfloat[ClueWeb09]{%
\includegraphics[width=0.49\textwidth]{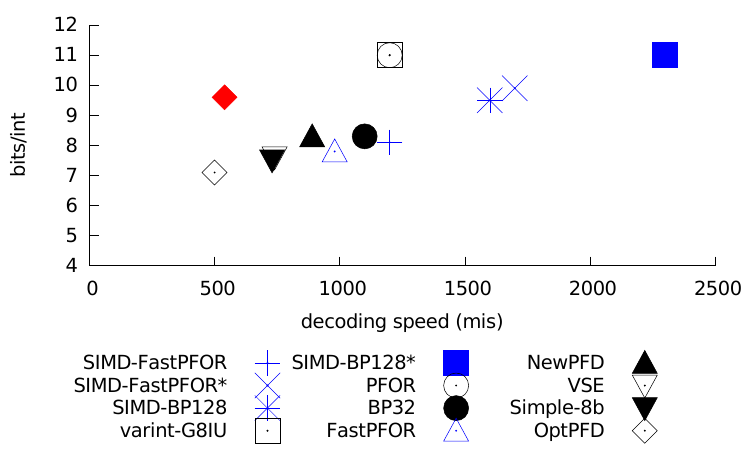}
}
\subfloat[GOV2]{%
\includegraphics[width=0.49\textwidth]{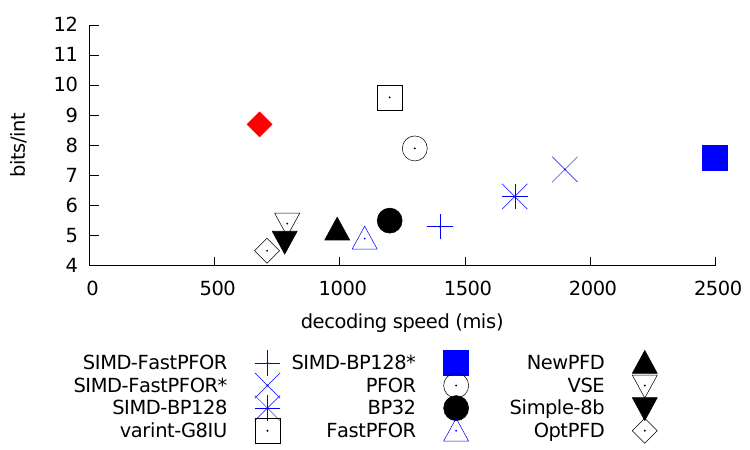}
}
\caption{\label{fig:scatter}Scatter plots comparing competitive schemes on decoding speed and bits per integer weighted based on AOL query logs. We use VSE as a shorthand for VSEncoding. For reference, Variable Byte is indicated as a red lozenge. The novel schemes (e.g., SIMD-BP128$^{\star}$) are identified with blue markers.}
\end{figure}

 For some applications, decoding speed and compression ratios are the most important metrics. 
  Whereas elsewhere we report the number of bits per integer $b$, we can easily compute the compression ratio as $32/b$.
  We plot both metrics for
  some competitive schemes (see Fig.~\ref{fig:scatter}). These plots suggest that the most competitive schemes are SIMD-BP128$^{\star}$, 
SIMD-FastPFOR$^{\star}$, \mbox{SIMD-BP128}, \mbox{SIMD-FastPFOR}, SimplePFOR, FastPFOR, Simple-8b, and OptPFD depending on how much compression is desired. Fig.~\ref{fig:scatter} also shows 
that to achieve decoding speeds higher than 1300\,mis, we must choose between SIMD-BP128, SIMD-FastPFOR$^{\star}$, and \mbox{SIMD-BP128$^{\star}$}.

Few research papers report encoding speed. Yet
we find large differences: for example, VSEncoding and OptPFD are two orders of magnitude slower during
encoding than our fastest schemes. 
If the compressed arrays are written to slow disks in a batch mode, such differences might be of little practical significance. However, for memory-based databases and network applications, 
slow encoding speeds could be a concern.
For example, the output of a query might need to be
compressed or we might need
to index the data in real time~\cite{netfli}.
 Our SIMD-BP128 and SIMD-BP128$^{\star}$ schemes have especially fast encoding.

Similarly to previous work \cite{Stepanov:2011:SDP:2063576.2063627,Silvestri:2010:VEC:1871437.1871592},
in Table~\ref{table:unweigthedaggrealistics} we report unweighted averages.
The unweighted speed aggregates are equivalent to computing the average speed over all arrays---irrespective of their lengths.
From the distribution of posting size logarithms in Fig.~\ref{fig:datadist}, 
one may conclude that weighted results should be  similar to unweighted ones.
These observations are supported by data in Table~\ref{table:unweigthedaggrealistics}:
the decoding speeds and compression ratios for both aggregation approaches differ by less than 15\% with the weighted results presented in Table~\ref{table:aggrealistics}. 

We can compare the number of bits per integer in Table~\ref{table:unweigthedaggrealistics} with an information-theoretic limit. Indeed,
the Shannon entropy for the deltas of ClueWeb09 is 5.5\,bits/int whereas it is 3.6 for GOV2. Hence, OptPFD is within 16\% of the entropy on ClueWeb09 whereas it is within 22\% of the entropy on GOV2.  Meanwhile, the faster SIMD-FastPFOR is within 30\% and 40\% of the entropy for ClueWeb09 and GOV2. Our fastest scheme (SIMD-BP128$^{\star}$) compresses the deltas of GOV2 to twice the entropy. It does slightly better with ClueWeb09 ($1.8\times$).

\begin{table}
\caption{\label{table:unweigthedaggrealistics}Average  speeds  in millions of 32-bit integers per second and bits per integer over all arrays of two data sets. These averages are \textbf{not} weighted according to the AOL query logs.}
\centering
\subfloat[ClueWeb09]{%
\begin{tabular}{lrrc}
 & coding & decoding  & bits/int \\ \hline 
SIMD-BP128$^{\star}$   &      \textbf{1600} &\textbf{2400}& 9.7\\
SIMD-FastPFOR$^{\star}$&                340 &1700 &9.0\\
SIMD-BP128             &         1000 &1700& 8.7\\
varint-G8IU$^{\star}$   &            230& 1400 &12\\
SIMD-FastPFOR            &             260 &1300 &7.2\\
PFOR2008                 &           260 &1300 &9.6\\
PFOR                      &           330 &1300 &9.6\\
varint-G8IU                &              220& 1200 &10\\
BP32                   &           770& 1100 &7.5\\
SimplePFOR          &        250 &1000 &6.9\\
FastPFOR             &         240& 1000& 6.9\\
NewPFD        &110 &900& 7.4\\
VSEncoding     &                    11 &760& 6.9\\
Simple-8b       &               290 &750 &6.7\\
OptPFD        & 16& 530& \textbf{6.4}\\
Variable Byte      &             630& 600 &9.2\\
\end{tabular}
}
\subfloat[GOV2]{%
\begin{tabular}{lrrc}
 & coding & decoding  & bits/int \\ \hline 
    &   \textbf{1600} &\textbf{2500}& 7.4\\
       &     350 &1900& 6.9\\
  &1000 &1800 &6.1\\
          &                 250& 1500 &10\\
    &     290 &1400 &5.1\\
  &    250 &1300& 7.6\\
    &  310 &1300 &7.6\\
  &                  230 &1200& 9.6\\
  &        790 &1200 &5.3\\
      &         260 &1100 &4.7\\
  &             270 &1100 &4.8\\
   & 150 &990 &5.0\\
                  &            11& 790 &5.4\\
                 &              340& 780& 4.6\\
 &    25& 720 &\textbf{4.4}\\
  &                          750 &700  &8.6\\
\end{tabular}
}
\end{table}

\section{Discussion}
We find that binary packing is both fast and space efficient.
The vectorized binary packing (\mbox{SIMD-BP128$^{\star}$}) is our fastest scheme.
While it has a lesser compression efficiency compared to Simple-8b,   
it is more than 3~times faster during decoding.
Moreover, in the worst case, a slower binary packing scheme (BP32) incurred a cost of only about 1.2~bits per integer compared to  the patching scheme with the best compression ratio (OptPFD) while decoding nearly as fast (within 10\%) as the fastest patching scheme (PFOR). 

Yet only few authors considered binary packing schemes or its vectorized variants in the recent literature:
\begin{itemize}
\item  Delbru et al.~\cite{Delbru2011} reported good results with a binary packing scheme similar to our BP32: in their experiments, it surpassed Simple-8b as well as a patched scheme (PFOR2008). 
\item Anh and Moffat~\cite{1034897} also reported good results with a binary packing scheme: in their tests, it decoded at least 50\% faster  than either Simple-8b or PFOR2008. As a counterpart, they reported that their binary packing scheme had a poorer compression.
\item  Schlegel et al.~\cite{Schlegel:2010:FIC:1869389.1869394} proposed a scheme similar to SIMD-BP128. 
This scheme (called \mbox{$k$-gamma}) uses
a vertical data layout to store integers, like our SIMD-BP128 and SIMD-FastPFOR schemes.
It essentially applies binary packing to tiny groups of integers (at most 4 elements).
From our preliminary experiments, we learned that decoding integers in small groups is not efficient.
This is also supported by results of Schlegel et al.~\cite{Schlegel:2010:FIC:1869389.1869394}.
Their fastest decoding speed, which does not include writing back to RAM, is
only 1600\,mis (Core~i7-920, 2.67\,Ghz). 
\item Willhalm et al.~\cite{Willhalm:2009:SUF:1687627.1687671} used a vectorized binary packing like our SIMD-BP128, but with a horizontal data layout instead of our vertical layout.
The decoding algorithm relies on the shuffle instruction \texttt{pshufb}. 
Our experimental results suggest that our approach based on a vertical layout might be preferable (see Fig.~\ref{fig:bitpackingcpp}): our implementation of bit unpacking over a vertical layout is sometimes between 50\% to  70\% faster than our reimplementation over a horizontal layout based on the work of Willhalm et al.~\cite{Willhalm:2009:SUF:1687627.1687671}. 

This performance comparison depends on the quality of our software.
Yet the speed of our
 reimplementation   is  comparable with the speed originally reported by Willhalm et al.~\cite[Fig.~11]{Willhalm:2009:SUF:1687627.1687671}: they report a speed of $\approx$3300\,mis with a bit width of 6. In contrast, using our implementation of their algorithms, we got a speed above 4800\,mis for the same bit width and a 20\% higher clock speed on a more recent CPU architecture. 

The approach described by Willhalm et al.\ might be more competitive on platforms  with instructions for simultaneously shifting several values by different offsets (e.g., the \texttt{vpsrld} AVX2 instruction). Indeed, this must be otherwise emulated by multiplications by powers of two followed by shifting.

 \end{itemize} 

Vectorized bit-packing schemes are  efficient: they encode/decode integers at speeds
of 4000--8500\,mis. Hence, the computation of deltas and prefix sums may become a major bottleneck.
This bottleneck can be removed through vectorization of these operations (though at expense of poorer compression ratios in our case).
We have not encountered this approach in the literature: 
perhaps, because for slower schemes the computation of the prefix sum accounts for a small fraction of total running time. 
In our implementation, to ease comparisons, we have separated differential decoding from
data decompression: an integrated approach could be up to twice as fast 
in some cases.
Moreover, we might be able improve the decoding speed and the compression ratios with 
better vectorized algorithms.
There might also be alternatives to data differencing, which also permit vectorization,
such as linear regression~\cite{Ao:2011:EPL:2002974.2002975}.

In our results, the original patched coding scheme (PFOR) is bested on all three metrics (compression ratio, coding and decoding speed) by a binary packing scheme (SIMD-BP128). Similarly, a more recent fast patching scheme (NewPFD) is generally bested by another binary packing scheme (BP32). Indeed, though the compression ratio of NewPFD is up to  6\% better on realistic data, NewPFD is at least 20\% slower than BP32.
Had we stopped our investigations there, we might have been tempted to conclude that patched coding is not a viable solution when decoding speed is the most important characteristic on desktop processors. 
However, we designed a new vectorized patching scheme \mbox{SIMD-FastPFOR}. 
It shows that patching remains a fruitful strategy even 
when SIMD instructions are used. 
Indeed, it is faster than the SIMD-based varint-G8IU  while providing a much better compression ratio (by at least 35\%).
In fact, on realistic data, SIMD-FastPFOR is better than BP32 on two key metrics: decoding speed and compression ratio (see Fig.~\ref{fig:scatter}).

In the future, we may expect increases in the arity of
SIMD operations supported by commodity CPUs (e.g., with AVX) as well as in memory speeds (e.g., with DDR4 SDRAM).
These future improvements could make our vectorized schemes even faster in comparison to their scalar counterparts.
However, an increase in arity means an increase in the minimum block size. Yet, when we increase the size of the blocks in binary packing, we also make them less space efficient in the presence of outlier values. Consider that BP32 is significantly more space efficient than \mbox{SIMD-BP128} (e.g., 5.5\,bits/int vs. 6.3\,bits/int on GOV2).

Thankfully,  the problem of outliers in large blocks can be solved through patching. 
Indeed, even though OptPFD uses the same block size as SIMD-BP128, it offers significantly better compression (4.5\,bits/int vs. 6.3\,bits/int on GOV2).
Thus, patching  may be more useful for future computers---capable of processing larger vectors---than for current ones.

While our work focused on decoding speed, 
there is promise in directly processing data while still in
compressed form, ideally by using vectorization~\cite{Willhalm:2009:SUF:1687627.1687671}. We expect that
conceptually simpler schemes (e.g., SIMD-BP128) might
have the advantage over relatively more sophisticated
alternatives (e.g., SIMD-FastPFOR) for this purpose.

Many of the fastest schemes use relatively large blocks
(128~integers) that are decoded all at once.
Yet not all queries require decoding the entire 
array. For example, consider the computation of intersections between sorted 
arrays. 
It is sometimes more efficient to use random
access, especially  when processing  arrays with vastly different lengths~\cite{Baeza-Yates:2005:EAF:2178997.2179000,Ding:2011:FSI:1938545.1938550,Vigna2013}. 
If the data is stored in relatively large compressed blocks (e.g., 128~integers with SIMD-BP128), the granularity of random
access might be reduced (e.g., when implementing a skip list). Hence, we may end up having to scan
many more integers than needed.
However, blocks of 128~integers might not necessarily be an impediment to 
good performance. 
Indeed,
Schlegel et al.~\cite{Schlegel2011} were able to
accelerate the computation of intersections by a factor of 5 with vectorization using blocks of up to
 65\,536~integers.
 
\section{Conclusion}

We have presented new schemes that are up to twice as fast as the previously best available schemes in the literature while offering competitive compression ratios and encoding speed.
This was achieved by vectorization of almost every step including
differential  decoding. To achieve both high speed and competitive compression ratios, we introduced a new patched scheme  that stores exceptions in a way that permits a vectorization (SIMD-FastPFOR).

In the future, we might seek to generalize our results over more varied architectures
as well as to provide a greater range of tradeoffs between speed and compression ratio. 
Indeed, most commodity processors support  vector processing (e.g., Intel, AMD, PowerPC, ARM). 
We might also want to consider adaptive schemes that compress more aggressively when the data is more compressible and optimize for speed otherwise. For example, one could use a scheme such as varint-G8IU
for less compressible arrays and SIMD-BP128 for the more compressible ones. 
One could also use workload-aware compression: frequently accessed arrays could be optimized for decoding speed whereas least frequently accessed data could be optimized for high compression efficiency.
Finally, we should consider more than just  32-bit integers. For example,
some popular  search
engines (e.g.,  Sphinx~\cite{aksyonoff2011introduction})  support 64-bit
document identifiers. We might consider an approach similar to   Schlegel et al.~\cite{Schlegel2011} who decompose arrays of 32-bit integers into blocks of 16-bit integers.

\ack Our varint-G8IU implementation  is based on code by M.~Caron.
V.~Volkov provided better loop unrolling for differential coding. P.~Bannister provided a fast algorithm to compute the maximum of the integer logarithm of an array of integers. 
 We are grateful to N.~Kurz for his insights and for his review of the manuscript.
We wish to thank the anonymous reviewers for their valuable comments. 

\bibliographystyle{wileyj}
\bibliography{./bib/longtitles,./bib/lemur}

\appendix

\section{Information-theoretic bound on binary packing}
\label{sec:itbinpack}
Consider arrays of  $n$~distinct sorted 32-bit integers. We can  compress the deltas computed from such arrays using binary packing as described in  \S~\ref{sec:binpacking} (see Fig.~\ref{fig:generalframework}).
We want to prove that such an approach is reasonably efficient.

There
are ${2^{32} \choose n}$~such arrays. Thus, by an 
information-theoretic argument, we need
at least $\log {2^{32} \choose n}$~bits to represent them.
By a well known inequality, we have that 
 $\log {2^{32} \choose n} \geq n \log \frac{2^{32}}{n} $.
 In effect, this means that we need at least $\log \frac{2^{32}}{n} $\,bits/int.
 
 Consider binary packing over blocks of $B$~integers: e.g., for BP32 we have $B=32$ and for \mbox{SIMD-BP128} we have $B=128$. 
 For simplicity, assume that the array length $n$ is divisible by $B$ and that $B$ is divisible by 32. Though our result also holds for vectorized differential coding (\S~\ref{sec:fast-delta}), assume that we use the common version of differential coding before applying binary packing. That is, if the original array is $x_1, x_2, x_3, \ldots$ ($x_i>x_{i-1}$ for all $i>1$), we compress the integers $x_1, x_2-x_1, x_3-x_2,\ldots$ using binary packing. 

For every block of $B$~integers, we have an overhead of $8$~bits to store the bit width~$b$.  This contributes $8 n /B $~bits to the total storage cost. The storage of any given block depends also on the bit width for this block. 
In turn, the bit width is bounded by the logarithm of the difference between the largest and the smallest element in the block.  If we write this difference for block $i$ as $\Delta_i$, 
the total storage cost in bits is  
  \begin{eqnarray*}
  \frac{8n}{B} + \sum_{i=1}^{n/B} B\lceil  \log (\Delta_i) \rceil & \le &   \frac{8 n}{B} + n + B \log \left (\prod_{i=1}^{n/B} \Delta_i \right).
  \end{eqnarray*}
  Because $\sum_{i=1}^{n/B} \Delta_i\leq 2^{32}$, 
we can show 
that the cost is maximized when $\Delta_i = 2^{32} B/n$.  
Thus, we have that the \emph{total} cost in bits is smaller than 
 \begin{eqnarray*}
  \frac{8 n}{B} +n+ B \log \left (\prod_{i=1}^{n/B} \frac{2^{32} B}{n} \right )    
  & = & 
  \frac{8 n}{B}+n + B \log \left ( \frac{2^{32} B}{n} \right )^{n/B}    
  \\
  & = &  \frac{8 n}{B} +n+ n \log  \frac{2^{32} B}{n}    ,\\
  \end{eqnarray*}
 which is equivalent to $8/B + 1+\log B+  \log  \frac{2^{32} }{n} $\,bits/int. 
Hence, in the worst case, binary packing is suboptimal by $8/B+1+\log B$\,bits/int.
 Therefore, we can show that BP32 is 2-optimal for arrays of length less than $2^{25}$~integers: its storage cost is no more than twice the information-theoretic limit. We also have that SIMD-BP128 is 2-optimal for arrays of length $2^{23}$~or less.

\end{document}